\documentclass[11pt,a4paper]{article}

\usepackage{jheppub}
\usepackage{xspace}
\usepackage{bbm}
\usepackage[normal]{caption}

\newcommand{\Real}{\ensuremath{\mathbb{R}}\xspace}

\newcommand{\Integer}{\mathbb{N}}

\newcommand{\op}{\ensuremath{\mathcal{O}}\xspace}

\newcommand{\mM}{\mathcal{M}}
\newcommand{\mE}{\mathcal{E}}

\newcommand{\mH}{\mathcal{H}}
\newcommand{\mA}{\mathcal{A}}
\newcommand{\mT}{\mathcal{T}}
\newcommand{\bH}{\mathbb{H}}
\newcommand{\Lie}{\pounds\hspace{-1.5pt}}

\newcommand{\hf}{\frac{1}{2}}

\newcommand{\qt}{\frac{1}{4}}

\let\a=\alpha  \let\g=\gamma  \let\e=\epsilon
    
\let\l=\lambda  
 \let\r=\rho
\let\s=\sigma    

  \let\D=\Delta

\let\del=\partial

\let\ve=\varepsilon

\let\na=\nabla

\newcommand{\be}{\begin{equation}}
\newcommand{\ee}{\end{equation}}
\newcommand{\bea}{\begin{eqnarray}}
\newcommand{\eea}{\end{eqnarray}}
\def\ba{\begin{array}}
\def\ea{\end{array}}

\def\del{\partial}

\def\ie{{\it i.e.\ }}

\def\eg{{\it e.g.\ }}

\def\oD{\,^{[0]}\hspace{-2pt}D}
\def\cov{\,^{(0)}\hspace{-1.2pt}\na}

\def\<{ \langle }
\def\>{ \rangle }

\def\({ \left( }
\def\){ \right) }

\newcommand{\Lra}{\Leftrightarrow}

\newcommand{\Tr}{{\rm Tr} }

\newcommand{\n}{\noindent}
\newcommand{\nn}{\nonumber}

\title{Holographic Reconstruction and Renormalization in Asymptotically Ricci-flat Spacetimes}

\author[a]{R. N. Caldeira Costa,}
\date{January 2012}

 \affiliation[a]{University of Amsterdam, Institute for Theoretical Physics,\\
 Science Park 904, Postbus 94485, 1090 GL Amsterdam, The Netherlands}

\emailAdd{R.N.Caldeira-Costa@uva.nl}

\abstract{\ In this work we elaborate on an extension of the AdS/CFT framework to a subclass of gravitational theories with vanishing cosmological constant. By building on earlier ideas, we construct a correspondence between Ricci-flat spacetimes admitting asymptotically hyperbolic hypersurfaces and a family of conformal field theories on a codimension two manifold at null infinity. By truncating the gravity theory to the pure gravitational sector, we find the most general spacetime asymptotics, renormalize the gravitational action, reproduce the holographic stress tensors and Ward identities of the family of CFTs and show how the asymptotics is mapped to and reconstructed from conformal field theory data. In even dimensions, the holographic Weyl anomalies identify the bulk time coordinate with the spectrum of central charges with characteristic length the bulk Planck length. 
Consistency with locality in the bulk time direction requires a notion of locality in this spectrum.
}

\begin{document}
\maketitle
\flushbottom

\newpage

\section{Introduction}

\qquad A holographic description of gravitational theories by lower dimensional quantum field theories without gravity has been the subject of extensive work over the last years. Such a correspondence has been successfully achieved in the context of string theory in asymptotically locally AdS manifolds by the AdS/CFT duality. According to this framework in its most common usage, weakly coupled gravitational theories on conformally compact, asymptotically Einstein manifolds with a negative cosmological constant can be reconstructed holographically from conformal field theory data on the conformal boundary of the manifolds. Reciprocally, the set of correlators and Ward identities defining a strongly interacting CFT\footnote{ 
More precisely, the dual quantum field theory need only to contain a UV-fixed point of the renormalization group.\\[-5pt]} 
can be reproduced from the asymptotic structure and regularity conditions of such a gravitational theory in one dimension higher. Essential ingredients in the recipe are the identification of the supergravity generating functional with that of the dual quantum field theory, together with the identification of the boundary configurations of bulk fields with the sources for gauge-invariant operators in the field theory.\\

The extension of the prescriptions defining AdS/CFT to gravitational theories with vanishing cosmological constant has proved much harder to accomplish in part due to the null nature of the conformal boundary of asymptotically Ricci-flat spacetimes. An impulsive application of the recipe would require a dual quantum field theory on a degenerate manifold \cite{witten-strings:1998} and in this case, the reconstruction of one side of the duality out of data from the other with the standard technology developed for AdS/CFT has remained unsurprisingly elusive. Several attempts have dropped the requirement for a conformal boundary and analysed instead the consequences of a possible dual description at spacelike infinity \cite{Kraus:1999di,Mann:2005yr,Marolf:2006bk,Papadimitriou:2010as}. An essential problem with this approach regards to the asymptotic structure of the bulk equations together with the divergences of the gravitational action. An asymptotic analysis of the Einstein and matter equations in this limit reveals that, unlike the case of a non-vanishing cosmological constant, the expansion of bulk fields near spatial infinity is not uniquely determined by the equations. Furthermore, if one imposes a given power expansion, the equations become differential, as opposed to algebraic, for the coefficients in the asymptotics. Such coefficients are therefore non-local with respect to each other which implies the inexistence of a set of local counterterms in the holographic renormalization programme that removes the divergences for any solution \cite{Solodukhin:1999zr,deHaro:2000wj,Skenderis:2002wp,Papadimitriou:2010as}. A different recent attempt \cite{Bredberg:2011jq,Compere:2011dx,Compere:2012mt} focuses on subregions of some Ricci-flat spacetime bounded by timelike hypersurfaces and reconstructs the metric of such a region from data belonging to a relativistic fluid describing the hydrodynamic regime of some QFT on the timelike boundary.\\

If one insists, on the other hand, on a notion of conformal boundary, it seems possible to avoid the technical problems of a dual description at null infinity along the lines of AdS/CFT by relaxing the codimension one condition in the prescription while keeping the remaining features and attempting to establish a duality in two dimensions less between a gravitational theory in an asymptotically Ricci-flat spacetime and a quantum field theory on a codimension two submanifold at null infinity.\footnote{ 
Such approach is closer in spirit to the conjectured dualities between four dimensional black holes and two dimensional conformal field theories at the horizons. It does not seem unlikely that putative holographic descriptions of Ricci-flat spacetimes and of black holes in a fashion similar to AdS/CFT share common features and a similar mechanics, since both geometries are bounded by null surfaces.\\[-5pt]}
The main problem with this approach lies in the reconstruction of two bulk dimensions, \ie the evolution of bulk fields along two extra dimensions, from quantum field theory data, which in addition cannot be both simultaneously spacelike. In AdS/CFT, the conformal boundary of the asymptotically Einstein spaces is timelike and any quantum field theory at the boundary need only to contain enough information to allow for the reconstruction of one spatial (radial) dimension. Such information is indeed captured  in the dynamics and kinematic constraints of the boundary theory from which the radial evolution of bulk fields, their spacetime dependence along the extra radial direction, can be reconstructed. In our case, on the other hand, the null nature of the conformal boundary requires the behaviour of bulk fields along the timelike direction to be captured as well by the dual field theory. Time evolution in the bulk must therefore be of central importance in such a holographic description of Ricci-flat spacetimes.\footnote{ 
It should be stressed that Ricci-flat spaces with Euclidean signature cannot have a conformal boundary. It is simple to show that the Ricci scalar of conformally compact Riemannian manifolds cannot vanish asymptotically, hence conformal compactness necessarily requires the Ricci-flat spaces to be Lorentzian, which also follows from the fact that the conformal boundary must be null. This implies in particular that one cannot have simultaneously a static conformal embedding and a time-independent defining function, otherwise a simple Wick rotation would violate the above statement. See appendix \ref{appA} for further details.\\[-5pt]}\\

In the work initially developed by de Boer and Solodukhin \cite{deBoer:2003vf,Solodukhin:2004gs} it was proposed that fields on {\it (d+2)}--dimensional Minkowski space could be reconstructed from conformal field theory data on a {\it d}--dimensional conformal manifold representing the boundary of the Minkowski lightcone. The key observations are a) that the interior of the lightcone is naturally foliated by conformally compact, hyperbolic (or Euclidean AdS) hypersurfaces whose boundaries all degenerate to the boundary of the lightcone and b) that the isometry group of (an asymptotically) Minkowski spacetime contains a subgroup that acts on such boundary as the conformal group. Since each leaf of the foliation admits a holographic description in terms of a conventional Euclidean {\it d}--dimensional CFT on its conformal boundary, the authors asked whether the interior of the lightcone could be described holographically in terms of a family of CFTs on this codimension two submanifold to which all boundaries converge. The problem raised by this approach regards to the reconstruction of the bulk timelike dimension from field theory data. The AdS/CFT dictionary allows one to reconstruct the radial evolution of bulk fields on each slice through the dynamics of the respective dual quantum field theory. The authors then left open the non-trivial possibility that evolution along the extra timelike direction defined by, and orthogonal to, the foliation could also be reconstructed from the infinite set of CFTs that reside on the boundary of the lightcone. Although the AdS/CFT duality guarantees that each CFT encodes radial evolution along each slice, it does not necessarily imply the family of CFTs should encode time evolution orthogonal to the slices.\\

The purpose of this work is to elaborate on such proposal and provide evidence supporting the conjecture that fields on a specific class of asymptotically Ricci-flat spacetimes can be reconstructed out of conformal field theory data on a codimension two conformal manifold representing the boundary of a null surface in the bulk that extends to null infinity. We will do so in the case of pure gravity\footnote{ 
Recall that in AdS/CFT, pure gravity in the bulk is dually described by a conformal field theory with vanishing vacuum expectation values (vevs) and correlators of every gauge-invariant operator with the exception of the CFT energy tensor. This is indeed the picture that arises by working with the full supergravity action, performing the holographic computations and in the end setting the bulk matter to zero. The field theory holographically described by pure bulk gravity is therefore in a state in which no operator has dynamics but the energy tensor.}
by generalising the proposal of de Boer and Solodukhin to such class of spacetimes and showing that both radial and time evolution of the bulk spacetime metric can be reconstructed from CFT data, in particular from the conformal structure on such codimension two manifold and from the expectation values of a family of conformal field theory stress tensors. Our results will indeed be consistent with a dual description of the time evolution of the Ricci-flat metric by a family of conformal field theories. The procedure will follow the standard AdS/CFT programme at the full non-linear level for our class of spacetimes by finding the most general spacetime asymptotics towards the conformal boundary, holographically renormalizing the gravitational action, computing the expectation values and Ward identities of the stress tensors of the field theories and mapping these to the data necessary to the reconstruction of the bulk metric. The results obtained for the holographic Weyl anomalies in even dimensions then imply that, for each CFT, the bulk timelike coordinate plays the role of the CFT central charge(s), with the bulk Planck length as the characteristic length. We elaborate more on this aspect in section \ref{holog_foliation} and then mainly in \ref{holog_energy-tensor}.\\

In the next section we review the foliation of Minkowski space that motivates our framework and describe the generalisation to a specific class of asymptotically Ricci-flat manifolds. We then outline the approach taken to deducing the most general asymptotics of such spacetimes and which is based on the initial value formulation of general relativity. In section \ref{initial_value} we briefly review the latter formalism and apply it to our class of manifolds. Solving the equations of motion within such framework will allow us to obtain in a unique way the asymptotic behaviour of the metric for such spacetimes and to find its relation to the ambient metric of Fefferman and Graham \cite{FeffermanGraham}. In section \ref{renormalization} we renormalize holographically the gravitational action and compute the vacuum expectation values and Ward identities of the family of dual field theories. The last section represents a generalisation of the previous formalism. Our class of asymptotically Ricci-flat manifolds will be generalised further by including subleading corrections to the spacetime asymptotics. This will allow us to obtain different expectation values for different field theories in this family. In the appendix we provide a few definitions that are necessary to our formalism together with several technical results.

\section{Preliminaries}\label{sec1}
\subsection{Foliation of Minkowski space}

\begin{figure}[h!]
\centering
\includegraphics[width=4.0in]{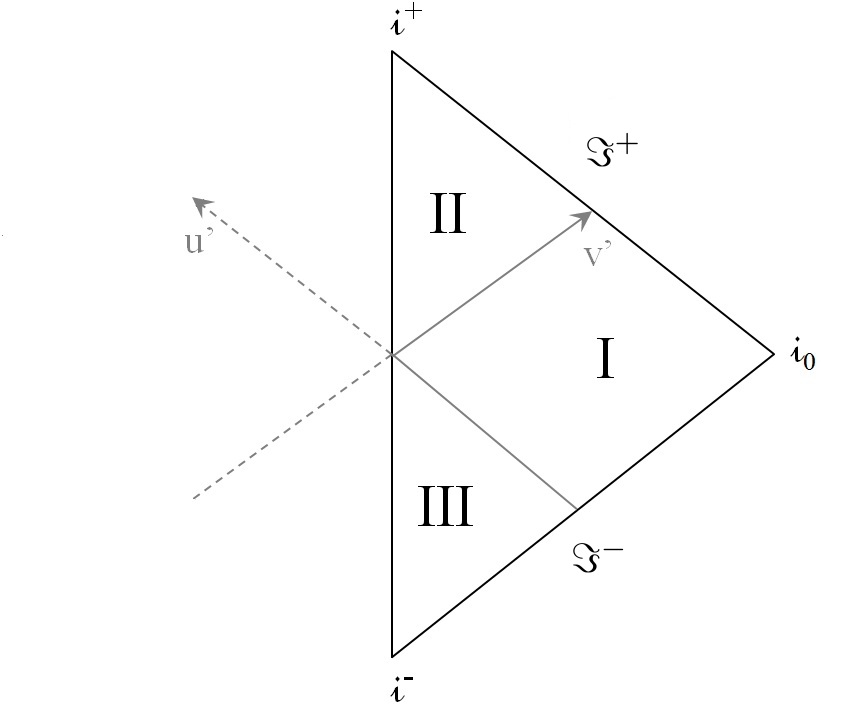}
\caption{\small{Penrose diagram for Minkowski space}}
\label{Mink_penrose}
\end{figure}

Let $(\mM,G_{\mu\nu})$ be {\it (d+2)}--dimensional Minkowski space. In spherical coordinates:
\be\label{mink}
ds^{2}_{d+2} = -dT^{\,2} + dr^{2} + r^{2}d\Omega_{d}^{2}\ .
\ee
Let one introduce null coordinates $\(v := T + r\, ,\, u := T - r\)$:
\be\label{mink_null}
ds^{2}_{d+2} = -dvdu + v^{2}\({1-u/v \over 2}\)^{2}d\Omega_{d}^{2}\ ,
\ee
such that infinity is represented by the union of the regions: \makebox{$\Im^{+} = \{v = +\infty\, , |u|<\infty\}$ ,} $\Im^{-} = \{u = -\infty\, , |v|< \infty\}$ , $i_{\pm} = \{ v = u = \pm \infty : |v-u| < \infty \}$ and: $i_{0} = \{ v = -u = +\infty : |v+u| < \infty \}$. To bring these regions to finite values of the coordinates, one introduces Penrose-type null coordinates $\(v' := \arctan v\, ,\, u' := \arctan u\)$:
\be
ds^{2}_{d+2} = {1 \over \cos^{2}v'\, \cos^{2}u'} \( -dv'\,du' + \qt \sin^{2}(v'-u')\,d\Omega_{d}^{2}\) := \r^{-2}(x)\, d\tilde{s}^{2}_{d+2}\ ,
\ee
where infinity is represented by the region where the defining function $\r(x):= \cos v'\, \cos u'$ vanishes. Since $r > 0$ in \eqref{mink}, then $v' > u'$ and the flattened Penrose diagram for the conformal embedding $(\tilde{\mM},\tilde{G}_{\mu\nu}=\r^{2}G_{\mu\nu})$ is given in figure \ref{Mink_penrose}, where each point represents a $S^{d}$ (with the exception of the corners $i_{\pm,0}$).\footnote{ More precisely, the conformal embedding is obtained from $\tilde{\mM}$ by deleting the corners $i_{\pm,0}$. Notice that $d\rho = 0$ and $\tilde{G}$ is degenerate in those regions, hence the triple $(\tilde{\mM},\tilde{G},\rho)$ does not represent an asymptote unless the corners are removed from the conformal embedding. See appendix \ref{appA} for further details.}

\begin{figure}
\centering
\includegraphics[width=6.0in]{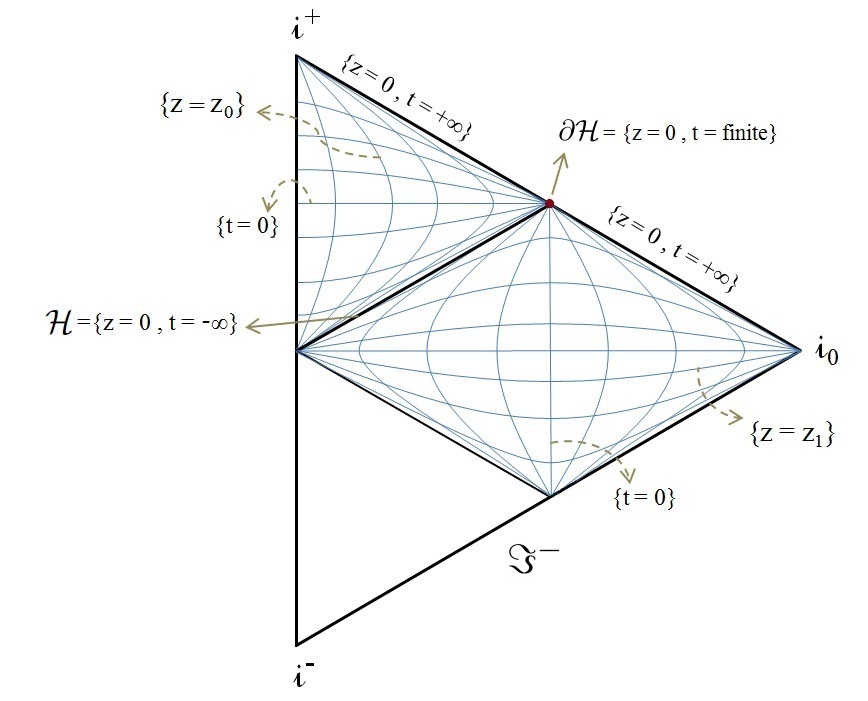}
\caption{\small{Foliation of Minkowski space}}
\label{Mink_foliation}
\end{figure}

Let one now return to the non-compact coordinate system \eqref{mink_null}. We are interested in looking at the region near $\Im^{+}$ in two particular charts. We introduce Rindler-type coordinates $(z,t)$ defined as: $\( z\,e^{t}:= u \ ,\ z^{-1}e^{t} := v \)$ such that:
\be\label{mink_II}
ds^{2}_{\,\text{II}} = {e^{2t} \over z^{2}}\,\Bigg( dz^{2} - z^{2}dt^{2} + \({1-z^{2} \over 2}\)^{2} d\Omega_{d}^{2} \Bigg)\ .
\ee
This coordinate system covers only region II of Minkowski space, where $v,u>0$. Notice that, due to the coordinate singularities at $z=\pm 1,0$, these coordinates are only defined in the interval: $z \in\, ]0,1[$. In order to cover region I (only) we analytically continue $t$ and $z$ to complex values: $\(z \to i z\, , t \to t - i\pi/2 \)$, which is equivalent to defining: $\( z\,e^{t}:= u \ ,\ z^{-1}e^{t} := -v \)$ in the original coordinate system \eqref{mink_null}:
\be\label{mink_I}
ds^{2}_{\,\text{I}} = {e^{2t} \over z^{2}}\,\Bigg( -dz^{2} + z^{2}dt^{2} + \({1+z^{2} \over 2}\)^{2} d\Omega_{d}^{2} \Bigg)\ .
\ee
In this case, the coordinate $z$ is defined in the interval $z \in\, ]-\infty,0[$. The Penrose diagram in each region in the new coordinates is given in figure \ref{Mink_foliation}, 
with $z_{0}$ and $z_{1}$ a positive and a negative constant, respectively. It is relevant to notice that worldlines in region II orthogonal to the surfaces of constant $t$, \ie with tangent vector the unit normal $n = e^{-t} \partial_{t}$ to such surfaces (also called Eulerian worldlines), define geodesic observers. In Minkowski coordinates \eqref{mink}:
\be\label{mink_geod}
n\ =\ {1 \over \sqrt{1-V^{2}}}\ \Big( \partial_{T} + V\, \partial_{r} \Big)\ :\ V = dr/dT\ ,
\ee
which is the standard velocity of inertial particles in radial motion, where the constant relative velocity $V$ is related to the (constant) coordinate $z$ as: $V = {1 - z^{2} \over 1+ z^{2}}$. This feature will be revisited later in section \ref{renormalization}.

Future null infinity in each region is given by: $\Im^{+} =$ $\{z=0 : t \neq - \infty\}$ and the defining function in the new coordinates becomes:
\be
\rho(x)^{2} = {z^{2} \over e^{2t}}\, \( 1 + 2\,z^{2}\cosh(2t) + z^{4} \)\ .
\ee
This implies in particular that the conformal embedding for region II approaches Rindler space times $S^{d}$ as $z \to 0$:
\be
d\tilde{s}^{2}_{\ \text{II}} \sim dz^{2} - z^{2}dt^{2} + \qt d\Omega_{d}^{2}\ ,
\ee
with Rindler horizon $\{z=0\}$ represented by the union $\Im^{+} \cup \mH$ in region II with bifurcation point $\del \mH \subset \Im^{+}$. The null surface \makebox{$\mH = \{ u = 0\, , 0<v<\infty \}$} $= \{ z = 0\, ,\, t = -\infty : 0< ze^{-t} < \infty \}$ represents the boundary of the past domain of dependence of any partial Cauchy surface in region II and is therefore the past Cauchy horizon of this region. This horizon defines the future lightcone of Minkowski space with respect to an inertial observer at the origin $\{ r = 0 , T = 0 \}$.\\

\subsection{Holographic foliation and generalisation}\label{holog_foliation}

\qquad As we move into region II from region I, the timelike surfaces of constant $t$ asymptote to $\mH$ and become spacelike as we cross it. These surfaces are hyperbolic manifolds ($\bH_{d+1}$) in region II and de Sitter in region I, as well as conformally compact. The (future) conformal boundary of each such surface converges to a common region $\del \mH = \{ u = 0\, , v = +\infty \} = \{z = 0\, : z\,e^{t} = 0 = z\,e^{-t}\}$ representing the boundary of $\mH$. Notice that the coordinate $t$ degenerates on $\del \mH$ where it can assume any value. According to the AdS/CFT correspondence, each $\bH_{d+1}$ surface admits a dual description in terms of a {\it d}--dimensional Euclidean conformal field theory on its conformal boundary at $\del \mH$ and in particular, fields on each surface (including the induced metric) can be reconstructed out of CFT data at $\del \mH$. Given a family of CFTs at this boundary, one is therefore able to reconstruct a collection of hyperbolic hypersurfaces and their fields, but not necessarily able to reconstruct fields in the flat spacetime foliated by such surfaces. The AdS/CFT dictionary allows one to reconstruct from CFT data the evolution along the $z$-direction of the pullback of bulk fields to each slice, but it does not determine the evolution of bulk fields along the time direction orthogonal to the slices. In order to obtain a dual description of such behaviour, one needs to find asymptotically (near $\del \mH$) the most general time evolution of a given class of bulk fields and then to determine how the bulk data necessary to the reconstruction of such evolution is mapped to data in the family of CFTs.\\

Our main goal will be to obtain such asymptotics for the spacetime metric and to show that it is possible to reconstruct its evolution near $\del \mH$ from data belonging to a family of conformal field theories at this boundary, in particular from the conformal structure at $\del \mH$ and from the expectation values of the stress tensors of each field theory. We will also find in the case of even dimensions that the bulk time coordinate $t$ defining the leaves of the foliation essentially plays the role of the central charges of the CFTs. This feature is not at all surprising: each leaf of the foliation dually described by a unique CFT is uniquely defined by a hypersurface condition $\{t = constant\}$ and the time dependence of the metric (in our gauge \eqref{mink_II} the factor $e^{t}$) represents, on each slice, the AdS radius $\ell$ of the hyperbolic hypersurface, which from AdS/CFT is mapped to the central charges of the respective dual field theory \cite{Brown:1986nw,Henningson:1998gx,Henningson:1998ey}.\\

In order to reconstruct holographically the spacetime metric, we need to generalise the procedure developed in the previous section to a larger class of spacetimes. We therefore generalise Minkowski space by any conformally compact, asymptotically Ricci-flat manifold that admits an asymptotically hyperbolic hypersurface of constant mean curvature. In section \ref{horizon} we will analyse how the mean curvature condition may be relaxed. Since the conformal boundary of such manifolds is necessarily null, the hypersurface must extend to null infinity (as opposed to a spatial infinity). Such hypersurfaces are called hyperboloidal (see \eg \cite{Friedrich:1983,Zenginoglu:2007jw,frauendiener:1993,Frauendiener:1998}) and represent the natural generalisation of the hyperbolic leaves in the previous foliation of Minkowski space. Our starting point will be such initial hypersurface and we will then generate the Ricci-flat embedding near the hypersurface by time evolving it as follows.

If a Ricci-flat space admits a spacelike hypersurface, then the Ricci-flat neighbourhood of the surface can be identified with its ``time evolution'' in the ADM sense \cite{wald}, \ie the vacuum Einstein equations in this region are completely equivalent to the (gauge-fixed) Gauss-Codazzi equations for the surface, also called ADM, or initial value equations. The solution to the latter equations represents the time evolution of the induced metric of the hypersurface. Given such solution, together with the lapse function and the shift vector, one can then construct the most general metric for the Ricci-flat embedding near the surface as explained in the next section. The embedding will then be foliated by the different instances of the time-evolved hypersurface.

\begin{figure}[h!]
\centering
\includegraphics[width=2.5in]{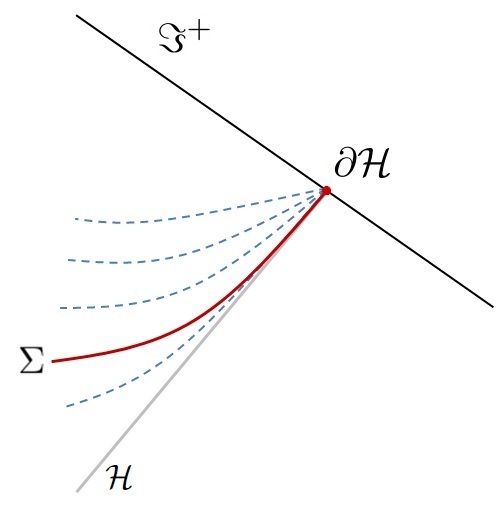}
\caption{\small{Ricci-flat embedding near the initial hypersurface $\Sigma$. The dashed lines represent the different time slices obtained by time evolving $\Sigma$ and the generated embedding represents the region between some initial and final slices (which can be taken to be infinitesimally close to $\mH$ and to $\Im^{+}$, respectively).}}
\label{hypersurface}
\end{figure}

If the initial surface $\Sigma$ is in particular asymptotically hyperbolic of constant mean curvature, then it is possible to solve asymptotically the Gauss-Codazzi equations and therefore to obtain the most general asymptotics of such Ricci-flat embeddings. Furthermore, since $\Sigma$ extends to null infinity $\Im^{+}$ by definition, the region $\del \mH$ in our previous case of Minkowski space will now represent that where $\Sigma$ intersects $\Im^{+}$ and we will then verify that the boundary of each time slice converges to $\del \mH$. Moreover, $\del \mH$ also represents the intersection of some null surface with null infinity. Such null surface will then be the past Cauchy horizon of the (generated) embedding and represents the generalisation of the null horizon $\mH$ that we found in the case of Minkowski space (see figure \ref{hypersurface}. Notice that the region to the right of $\mH$ contains future-inextendible causal curves that do not intersect $\Sigma$).

\subsection{Exterior of the future lightcone}

\qquad So far we have restricted our attention to a generalisation of region II of Minkowski space, the future domain of dependence of $\mH$. Region I represents the exterior of the lightcone in which the leaves of the foliation are de Sitter hypersurfaces with future conformal boundary converging to $\del \mH$, whereas region III is the interior of the past lightcone with a hyperbolic foliation as in region II. In the particular case of Minkowski space, all the results obtained for region II may be extended to region I or III by analytic continuation: $\(z \to i z\, , t \to t - i\pi/2 \)$, or radial and time reversal: $\(z \to -z\, , t \to -t\)$, respectively. Throughout this work we will restrict our attention to a single Ricci-flat region generated by time evolving an initial Cauchy surface as explained above and which will be realised in the next section. The generalisation of regions I and III can then be obtained from such solution by the above continuations.\footnote{
It should be emphasized, however, that the generalisation of region I would involve generating a Ricci-flat region by evolving a timelike surface along a spatial direction. While the above analytic continuation should result in the appropriate generalisation of region I, it must be emphasized that the initial data surface is not Cauchy in such case. This implies that the uniqueness property of solutions to the initial value problem in the case of Cauchy initial data surfaces proved by Choquet-Bruhat and reviewed in the next section do not necessarily carry over to this case, hence the unique spacetime asymptotics that we will later obtain may no longer be unique when analytically continued. See also appendix A of \cite{Skenderis:2002wp} on the analytic continuation of solutions in the context of AdS to dS/CFT.}

In the next section we briefly review the initial value formulation of relativity which will allow us to generate the Ricci-flat embeddings of initial data hypersurfaces. We will then use the existence and uniqueness properties of solutions to the Cauchy problem in order to obtain the spacetime asymptotics.

\section{Spacetime asymptotics}\label{initial_value}

\subsection{Initial value formulation}

\qquad Since we are only interested in (asymptotically) Ricci-flat embeddings, for the sake of simplicity we will focus on the vacuum version of the initial value formulation. A recent review of the subject can be found in \cite{Gourgoulhon:2007}.\\

An initial data set is a triple $\( \Sigma, h_{[0]}, K_{[0]} \)$ consisting of a Riemannian manifold $\Sigma$ with a positive-definite metric $h_{[0]ab}$ and a symmetric tensor field $K_{[0]ab}$ satisfying the {\it constraint equations}:
\bea
R[h_{[0]}] + (K_{[0]a}^{a})^{2} - K_{[0]ab}K_{[0]}^{ab} &=& 0 \quad ,\label{HamiltC}\\[10pt]
\oD_{b} K^{b}_{[0]a} - \oD_{a} K^{b}_{[0]b} &=& 0 \quad ,\label{DiffeoC}
\eea
where $R_{ab}[h_{[0]}]$ and $\oD_{a}$ are the Ricci tensor and the covariant derivative associated to $h_{[0]}$. Given an initial data set, one introduces a flow parameter $t$ called the development time and evolves the initial data in time by specifying a scalar and vector fields $N$ and $A^{a}$ and solving the first order differential equations:
\bea
&&\partial_{t} h_{ab}\ =\ \Lie_{A} h_{ab} + 2 N K_{ab} \quad ,\label{evEq}\\[5pt]
&&\partial_{t} K_{ab}\ =\ \Lie_{A} K_{ab} + D_{a}D_{b} N - N \Big( R_{ab}[h] - 2K_{\, a}^{c}K_{cb} + K K_{ab} \Big)\ ,\label{dynEq}
\eea
subject to the initial value conditions:
\be\label{i_values}
h_{ab}\big|_{t=0} = h_{[0]ab}\quad ,\quad  K_{ab}\big|_{t=0} = K_{[0]ab}\ ,
\ee
with $D_{a}$ the covariant derivative with respect to $h_{ab}$. After a solution is found, one constructs the metric tensor $G_{\mu\nu}$ of a Lorentzian manifold $\mM = I \times \Sigma$, where $I \in \Real$ is the interval over which the time evolution is carried, according to the formula:
\bea\label{dev_met}
ds^{2} &=& G_{\mu\nu}dx^{\mu}dx^{\nu}\nn\\[5pt]
&=& -N^{2} dt^{2} + h_{ab} \left( dx^{a} + A^{a}dt \right) ( dx^{b} + A^{b} dt )\ .
\eea
The manifold $( \mM, G)$ is called the development of the Cauchy surface $\Sigma$ and will satisfy the vacuum Einstein equations. Notice that every Ricci-flat space can be generated in this way because the equations \eqref{HamiltC}--\eqref{i_values} simply represent the Gauss-Codazzi identities\footnote{
By virtue of the Bianchi identities, the constraint equations have the fundamental property that they hold for all $t$ if they hold at a given $t$. In the conventional approach to relativity, equation \eqref{evEq} represents the definition of extrinsic curvature $K_{\mu\nu} = 1/2 \Lie_{n}h_{\mu\nu}$ projected onto the surfaces of constant $t$.\\[-5pt]} 
for a Ricci-flat space cast as a Cauchy problem. The development represents an embedding of the initial Cauchy surface $\Sigma = \{t = 0\}$ and is foliated by a one-parameter family of spacelike surfaces $\Sigma_{t}$ representing the time evolution of $\Sigma$, each defined by the condition $t = constant$ with future-directed unit normal $n = -N dt$. The induced metric and extrinsic curvature of such surfaces are given by:
\be\label{extension}
h_{\mu\nu} = G_{\mu a}G_{\nu b} h^{ab} = G_{\mu\nu} + n_{\mu}n_{\nu}\quad ,\quad K_{\mu\nu} = G_{\mu a}G_{\nu b} K^{ab} \ ,
\ee
where $K^{ab} = (h^{-1}Kh^{-1})^{ab}$. Finally, a choice of {\it lapse function} $N$ and {\it shift vector} $A^{a}$ as above is called a gauge choice and it represents how one chooses to foliate $\mM$ and how to propagate the coordinate system of $\Sigma$ in $\mM$. In other words, it represents a choice of coordinates for $\mM$ near $\Sigma$ and one is free to choose a lapse and a shift without changing the physical spacetime in such neighbourhood.\\

Choquet-Bruhat showed \cite{Choquet:1952} that there always exists a solution to the initial value problem for smooth initial data and that such solution is unique in a neighbourhood of the initial Cauchy surface. Choquet-Bruhat and Geroch \cite{Choquet:1969} also showed global existence and uniqueness of a maximal development, the element in the set of solutions into which every other solution can be isometrically mapped. See the review in \cite{wald}. These properties are the main reason behind our choice of approach to finding the spacetime asymptotics. We will make use of the existence and uniqueness of a solution to the Cauchy problem in order to deduce the most general asymptotics of the developments of initial data sets that asymptote to hyperboloidal sets. The latter are defined as follows.\\

An asymptotically hyperboloidal initial data set \cite{Friedrich:1983,Friedrich:1987,LAndersson:1992,LAndersson:1993} (see also \cite{LAndersson:1996,MAnderson:2001}) is defined as any initial data set $(\Sigma,h_{[0]},K_{[0]})$ such that $(\Sigma,h_{[0]})$ is a conformally compact, asymptotically hyperbolic manifold\footnote{
See appendix \ref{appA} about the equivalence between such manifolds and conformally compact, asymptotically Einstein Riemannian manifolds of negative scalar curvature.}
 and $K_{[0]}$ asymptotically covariantly conserved.\\

This last condition is equivalent to $K_{[0]}$ being asymptotically equal to $h_{[0]}$ up to a proportionality constant. This can be proved as follows (we work in a sufficiently small neighbourhood of the conformal boundary). If $K_{[0]}$ is equal to $h_{[0]}$, then it is covariantly conserved. Reciprocally, if $K_{[0]}$ is covariantly conserved, then it follows from the diffeomorphism constraint equation \eqref{DiffeoC} that the trace of $K_{[0]}$ is constant. Now, since $\Sigma$ is Einstein with negative scalar curvature, its Ricci tensor is proportional to the metric:
\be\label{Ricci_ell}
R_{ab}[h_{[0]}] = -{d \over \ell^{2}}\, h_{[0]ab}\ ,
\ee
with $d+1$ the dimension of $\Sigma$ and $\ell$ a real constant. If the trace of $K_{[0]}$ vanishes, then the Hamiltonian constraint equation \eqref{HamiltC} cannot be satisfied because $K_{[0]ab}K_{[0]}^{ab} \geq 0$ ($h_{[0]}$ is positive definite and $K_{[0]}$ is symmetric and real, see below). Hence, the trace of $K_{[0]}$ must be a non-zero constant and it can then always be normalised (\eg by rescaling the metric) such that:
\be\label{K_ell}
\Tr[h_{[0]}^{-1}K_{[0]}] = \pm {d+1 \over \ell}\ .
\ee
If we then decompose $K_{[0]}$ in terms of the shear tensor $\sigma_{ab}$ and the mean curvature:
\be\label{shear}
K_{[0]ab} = \sigma_{ab} + {1 \over d+1} h_{[0]ab} \Tr[h_{[0]}^{-1}K_{[0]}]\ ,
\ee
it then follows from the Hamiltonian constraint equation \eqref{HamiltC} that: $\sigma_{ab}\sigma^{ab} = 0$. 
Now, since $h_{[0]ab}$ is positive definite and $\sigma_{ab}$ Hermitian, then there exists an invertible matrix $Q$ such that (see \eg \cite{lancaster}):
\be
Q^{\dagger}h_{[0]}Q = \mathbbm{1}\quad ,\quad  Q^{\dagger}\sigma Q = D\ ,
\ee
with $D = \text{diag}(\l_{1},..,\l_{d+1}) : \l_{i} \in \Real\,$ and where $\dagger$ denotes the Hermitian conjugate.
In this way:
\be
0 = \Tr[h_{[0]}^{-1}\sigma h_{[0]}^{-1}\sigma] = \Tr[Q Q^{\dagger}\, (Q^{\dagger})^{-1}D Q^{-1}\, Q Q^{\dagger}\, (Q^{\dagger})^{-1}D Q^{-1}] = \Tr[DD] = \sum \l_{i}^{2}\ .
\ee
Hence: $D = 0$, which implies that $\sigma_{ab}$ vanishes and therefore, by \eqref{shear}, that $K_{[0]}$ is equal to $h_{[0]}$ up to a proportionality constant.\footnote{ The vanishing of the shear can also be seen by finding coordinates at a given point such that $h_{[0]}$ is locally the Euclidean metric. Then, since $\s_{ab}$ is real and symmetric, it can be diagonalised at each point by an orthogonal matrix. Replacing both conditions in the trace equation $\s_{ab}\s^{ab}=0$ implies $\s_{ab} = 0$ locally and hence everywhere since the equation is tensorial.\\[-5pt] }\\

Since $K_{[0]}$ represents the extrinsic curvature of the initial data surface $(\Sigma,h_{[0]})$ in the embedding, and as remarked above the diffeomorphism constraint implies that $\Sigma$ is of constant mean curvature, then the developments of asymptotically hyperboloidal sets are the asymptotically Ricci-flat spacetimes that we have introduced in the previous section.\footnote{ 
There is a technical point regarding conformal compactness. If the embedding of some Cauchy surface is conformally compact, then it is possible to show that the surface is also conformally compact (see \eg section 2 of \cite{Frauendiener:1998}). The reciprocal, however, is not necessarily true. In the above, we have only demanded that the initial data surface be conformally compact, whereas in the previous section we required conformal compactness of the Ricci-flat embedding. This means that, after finding the solution to our initial value problem, we will have to verify that the embedding is indeed conformally compact.\\[-5pt]} 
In this way, by finding the unique solution to the initial value problem with such initial data sets, we are able to find the asymptotics of such embeddings. We will begin by developing exact hyperboloidal sets and in section \ref{horizon} allow deviations of $K_{[0]}$ away from its asymptotic value by considering arbitrary subleading contributions.

\subsection{Ricci-flat asymptotics}\label{regionII}

\qquad Let $(\Sigma,h_{[0]},K_{[0]})$ be an asymptotically hyperboloidal initial data set as introduced in the previous section, with $(\Sigma,h_{[0]})$ of dimension $d+1$ and normalised such that: $R_{ab}[h_{[0]}] = -d\, h_{[0]ab}$ and: $K_{[0]} = h_{[0]}$ in a sufficiently small neighbourhood of the conformal boundary of $\Sigma$. It was found in \cite{LeBrun,FeffermanGraham}, see also \cite{deHaro:2000xn}, that coordinates can be found in which the most general asymptotics of this initial Cauchy surface takes the form:
\be\label{FG}
ds^{2}_{d+1}\, =\, h_{[0]ab}dx^{a}dx^{b}\, \sim\, {1 \over z^{2}} \( dz^{2} + g_{ij}dx^{i}dx^{j} \)\ ,
\ee
with the conformal boundary $\del \Sigma =\{z=0 \}$ and where $g_{ij}(z,x)$ is given asymptotically by the power series:
\be\label{FG_2}
g_{ij}(z,x)\, =\, g_{(0)ij}(x) + z^{2}g_{(2)ij}(x) + ... + z^{d} g_{(d)ij}(x) + z^{d}\log z\, \tilde{g}_{(d)ij}(x) + \op(z^{>d})\ ,
\ee
where only even powers of $z$ arise below the order $z^{d}$. The {\it non-normalisable mode} $g_{(0)}$ is an arbitrary field and each coefficient $g_{(n < d)}$, as well as $\tilde{g}_{(d)}$, is a local functional of $g_{(0)}$. For odd values of $d$, or for $d=2$, the logarithmic term vanishes and for even values it is traceless and divergenceless (with respect to $g_{(0)}$). The {\it normalisable mode} $g_{(d)}$ is undetermined up to its trace and divergence, which vanish for odd values of $d$ and are a local functional of $g_{(0)}$ for even values. See \cite{deHaro:2000xn} for the explicit expressions of these functionals.\\

\begin{figure}
\centering
\includegraphics[width=3.2in]{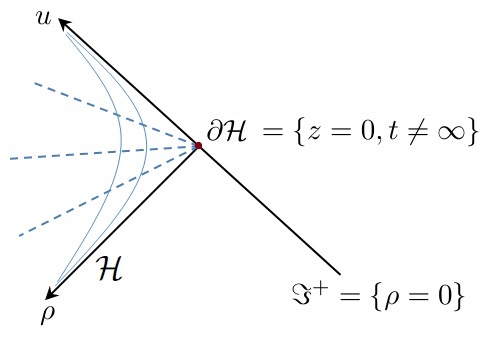}
\caption{\small{Conformal embedding near future null infinity. The dashed lines represent different surfaces of constant $t$, while the solid lines are surfaces of constant $z$. The null surface $\mH = \{z=0,t=-\infty\}$ is the past Cauchy horizon of the development.}}
\label{last_penrose}
\end{figure}

In order to evolve this initial data in time and generate a Ricci-flat development we need to prescribe a lapse function and a shift vector and we do so by choosing the geodesic normal gauge: $\( N=1, A^{a} = 0\)$, also called synchronous gauge, or Gaussian normal coordinates.\footnote{ Recall that any metric can be written in this gauge in a sufficiently small neighbourhood of a (non-null) hypersurface. Regarding the notation from the previous section, we will relabel our development time $t \to \hat{t}$.} In this gauge and sufficiently close to the Cauchy surface $\Sigma=\{\hat{t}=0\}$, the metric tensor \eqref{dev_met} of the development $\mM = I\times\Sigma$ takes the form:
\be\label{Gaussian_metric}
ds^{2}_{d+2} = -d\hat{t}^{\,2} + h_{ab}dx^{a}dx^{b}\ .
\ee
Since the constraint equations are trivially satisfied by the initial data, the only equation left to solve is the dynamical equation obtained by replacing equation \eqref{evEq} in \eqref{dynEq}:
\be
2 R_{ab}[h]  + \ddot{h}_{ab} + \hf \dot{h}_{ab} \Tr[h^{-1}\dot{h}] - \big( \dot{h}h^{-1}\dot{h} \big)_{ab} = 0\ ,
\ee
subject to the initial value conditions \eqref{i_values} and where $\dot{h}:=\partial_{\hat{t}}\,h$. The unique solution to this initial value problem is now very simple to find. One can easily begin by verifying that the ansatz:
\be\label{sol_1}
ds^{2}_{d+2} = -d\hat{t}^{\,2} + h_{ab}(\hat{t},x)dx^{a}dx^{b} = -d\hat{t}^{\,2} + \(1 + \hat{t}\,\)^{2} h_{[0]ab}(x)dx^{a}dx^{b}\ ,
\ee
is a solution to the dynamical equation, which is simply the well-known result that the Lorentzian cone of an Einstein metric is Ricci-flat. Furthermore, since the extrinsic curvature on the surfaces of constant $\hat{t}$ is given by: $K_{ab} = \hf \partial_{\hat{t}}\,h_{ab}$, one finds that:
\be
h_{ab} \big|_{\hat{t}=0} = h_{[0]ab}\quad ,\quad K_{ab}\big|_{\hat{t}=0} = h_{[0]ab}\ .
\ee
The metric \eqref{sol_1} is therefore a solution to this initial value problem and hence the unique solution. By performing the transformation of coordinates: $e^{t} := 1+\hat{t}$, our solution becomes:
\be\label{sol_2}
ds^{2}_{d+2} = e^{2t} \( -dt^{2} + h_{[0]ab}dx^{a}dx^{b} \) \ ,
\ee
where the Cauchy surface $\Sigma = \{ t = 0 \}$. Let us now denote by $\del \mH$ the region in the development\footnote{
Not strictly {\it in} the development, but in some appropriate conformal embedding.\\[-5pt]}
 described by the conformal boundary $\del \Sigma$  under time evolution over the interval $I \in \Real$ (\ie the portion of null infinity foliated by the leaves $t = constant$). Above we have found that, near $\del \Sigma = \{ z=0 \}$, $h_{[0]}$ takes the form \eqref{FG}. In this way, the development \eqref{sol_2} in a neighbourhood of $\del \mH$ takes the asymptotic form:\footnote{
In other words, this is the solution we would have found had we time evolved directly the asymptotic metric \eqref{FG}.\\[-5pt]}
\be\label{sol_3}
ds^{2}_{d+2} = {e^{2t} \over z^{2}}\,\Bigg( dz^{2} - z^{2}dt^{2} + g_{ij}dx^{i}dx^{j} \Bigg)\ , 
\ee
with $g_{ij}(z,x)$ given asymptotically by the expansion \eqref{FG_2}. In order to verify that the slices of constant time converge to $\del \Sigma$, and therefore that $\del \mH$ coincides with $\del \Sigma$, we bring in future null infinity to finite affine parameter distances by a suitable conformal compactification. We define coordinates $\( u := z\, e^{t}\ ,\ \rho := z\,e^{-t} \)$ such that:
\be\label{conf_comp}
ds^{2}_{d+2}\ =\ {1 \over \rho^{2}}\, \( d\rho du + g_{ij}dx^{i}dx^{j} \)\ ,
\ee
where:\footnote{
For $d=3$, we have instead $\op(\rho^{3/2})$. Recall that the logarithmic term in \eqref{FG_2} vanishes for $d=2,3$.\\[-5pt]}
$g_{ij} = g_{(0)ij} + \rho\,u\,g_{(2)ij} + \op(\rho^{2})$. Since $\tilde{G}_{\mu\nu}:= \rho^{2}G_{\mu\nu}$ is at least $C^{2}$ (or $C^{1}$ for $d=3$) and non-degenerate, then $(\tilde{\mM},\tilde{G})$ defines a conformal compactification with conformal boundary $\{\rho = 0\}$. The Penrose diagram near the boundary with the spatial coordinates $x^{i}$ suppressed (note that $g_{(0)} = g_{(0)}(x^{i})\,$) is then given by figure \ref{last_penrose}.

Returning to our asymptotic solution \eqref{sol_3}, if the Ricci-flat development is in particular Minkowski space as in \eqref{mink_II}, we have the expansion:
\bea
g_{ij} = g_{(0)ij} + z^{2} g_{(2)ij} + z^{4} g_{(4)ij}\ : \
\begin{cases}
g_{(0)ij}dx^{i}dx^{j} = \qt d\Omega_{d}^{2}\ ,\\
g_{(2)ij} = -2 g_{(0)ij}\ ,\\
g_{(4)ij} = g_{(0)ij}\ .
\end{cases}
\eea
More generally, in Minkowski space $g_{ij}$ can be expanded as:\footnote{
For the special case $d=2$ we have that $g_{(2)}-g_{(0)}\Tr[g_{(0)}^{-1}g_{(2)}]$ is the stress tensor of the Liouville field. See \cite{Skenderis:1999nb} for further details.}
\bea\label{Kostas}
g_{ij}(z,x) = g_{(0)ij} + z^{2} g_{(2)ij} + z^{4} g_{(4)ij}\ : \
\begin{cases}
g_{(2)ij} = -{1 \over d-2} \( R_{ij}[g_{(0)}] - {1 \over 2(d-1)} g_{(0)ij} R[g_{(0)}] \)\ ,\\[10pt]
g_{(4)ij} = \qt \( g_{(2)} g_{(0)}^{-1} g_{(2)} \)_{ij}\ ,
\end{cases}
\eea
where $g_{(0)}$ is any conformally flat metric. This can be easily seen by recalling that Minkowski space $\Real^{1,d+1}$ is the Lorentzian cone of the hyperbolic space $\bH_{d+1}$. It was found in \cite{Skenderis:1999nb} that $\bH_{d+1}$ in Poincar\'e coordinates is given by:
\be
ds^{2}_{d+1} = {1 \over z^{2}} \( dz^{2} + g_{ij}dx^{i}dx^{j} \)\ ,
\ee
where $g_{ij}$ is given by the expansion \eqref{Kostas} with $g_{(0)}$ any conformally flat metric. Hence, the respective cone is Minkowski. This implies in particular that the solution $g_{ij} = \delta_{ij}$ in \eqref{sol_3} is also Minkowski and the direct transformation of coordinates is given by:
\be
X_{0} := \displaystyle{e^{t} \over 2} \( z + \displaystyle{1 + \vec{x}^{2} \over z} \)\ ,\ Z := \displaystyle{e^{t} \over 2} \( z + \displaystyle{-1 + \vec{x}^{2} \over z} \)\ ,\ X^{i} := \displaystyle{x^{i} \over z}\, e^{t}\ ,
\ee
such that:
\bea\label{mink_poinc}
ds^{2}_{d+2} &=& {e^{2t} \over z^{2}}\,\Bigg( dz^{2} - z^{2}dt^{2} + dx^{i}dx^{i} \Bigg)\\
&=& -dX_{0}^{2} + dZ^{2} + dX^{i}dX^{i}\ . \nn
\eea
It is not difficult to show that the Penrose diagram for \eqref{mink_poinc} corresponds to that of region II in figure \ref{Mink_foliation} with the surfaces $t = constant$ and $z = constant$ still as in this diagram. Although the spacetimes \eqref{mink_II} and \eqref{mink_poinc} are both Minkowski and therefore diffeomorphic, we will find in the next section that they do not yield the same expectation values for the holographic stress tensors. These are computed from the (renormalized) gravitational action and vanish for the latter solution with future null infinity $\Real\times\Real^{d}$, whereas are non-vanishing for the former with null infinity $\Real\times S^{d}$ in the same fashion as in AdS holography \cite{deHaro:2000xn}. This is associated to the fact that the holographic renormalization scheme that we will employ later breaks invariance of the gravitational action with respect to bulk diffeomorphisms\footnote{
The bulk diffeomorphisms that preserve the asymptotic functional form \eqref{sol_3} of the metric contain a subgroup that generates conformal transformations at $\del \mH$. The proof of this fact is sketched in section 6 of \cite{deBoer:2003vf}, where our solution represents a particular case of the asymptotic metric analysed in this reference. A conformal transformation at $\del \mH$ is therefore realised in the bulk as an asymptotic ``isometry". See \cite{Imbimbo:1999bj} about the relation between bulk diffeomorphisms and conformal transformations at the boundary in the context of AdS/CFT.\\[-5pt]}
that result in a conformal transformation at $\del \mH$ and therefore spacetimes related by such transformations, such as \eqref{mink_II} and \eqref{mink_poinc}, will not necessarily result in the same renormalized expectation values as pointed out in \cite{deHaro:2000xn}.\\

So far we have found a coordinate system in which the most general asymptotics of our class of asymptotically Ricci-flat spacetimes assumes the form \eqref{sol_3}. It is now simple to see that our solution is diffeomorphic to the ambient metric of Fefferman and Graham \cite{FeffermanGraham} by recalling that the ambient construction represents the Lorentzian cone of an Einstein Riemannian manifold in coordinates adapted to the study of the past Cauchy horizon \cite{MAnderson:2001}. Indeed, by introducing coordinates $\( r := z^{2}, v := z^{-1}e^{t}\)$, our solution assumes the form:
\be\label{ambient}
ds^{2}_{d+2} = - r dv^{2} - v dv dr + v^{2} g_{ij}dx^{i}dx^{j}\ ,
\ee
which represents the ambient metric with $g_{ij}(r,x)$ expanded as in \eqref{FG_2} with $z = \sqrt{r}$. For $v$ finite, this represents an expansion away from the Cauchy horizon $\mH = \{ z=0, t=-\infty\} = \{r = 0, v \neq \infty\}$, whereas for $v = \infty$ it is an expansion away from $\del \mH = \{ z=0, t \neq \infty\} = \{r = 0, v = +\infty\}$. The coordinate system \eqref{ambient} is therefore well-adapted to the study of the former region, but unsuited to the study of the latter and the other way around with respect to \eqref{sol_3}. Since we are rather interested in the spacetime asymptotics, a correct choice of coordinates in our case is given by our original solution \eqref{sol_3}.\footnote{
Furthermore, notice that the metric \eqref{sol_3}, or \eqref{ambient}, represents the most general spacetime asymptotics near the boundary $\del \mH$ of $\mH$, but not necessarily the most general Ricci-flat metric near the entire $\mH$ unless we restrict further our class of spacetimes to those in which the metric in a neighbourhood of $\mH$ is identically equal to its asymptotic form. Such spacetimes are defined by the ambient metric defining conditions 1)--3) in section 2 of \cite{FeffermanGraham} (see also the conditions a)--d) in Problem 5.1 of this reference).}
As a matter of fact, rather than relying on a particular time coordinate, any coordinate system of the form:
\be\label{metric}
ds^{2}_{d+2} = - N(t)^{2}dt^{2} + {\beta(t)^{2} \over z^{2}} \( dz^{2} + g_{ij}dx^{i}dx^{j} \)\quad :\quad \beta(t) := \int dt N(t)\ \, ,\ \forall N(t)\neq 0\ ,
\ee
with $N(t)$ smooth, is suited to the study of the spacetime asymptotics and is related to our previous coordinates by: $ e^{t} \to \int dt N(t)$. We will find more useful to keep in this way the lapse function $N(t)$ arbitrary. In particular, we can now notice that the scalar field $\beta(t)$, which will play a central role in the remainder of this work, is a gauge invariant quantity, the coordinate invariant part of the lapse, and measures propertime distances along the so-called Eulerian worldlines, the timelike curves with tangent vector the future-directed unit normal $n = - Ndt = N^{-1}\partial_{t}$ to the constant time slices.\footnote{
For the interpretation of $\beta$ as a thermodynamic variable, see \cite{Brown:1992bq,Brown:1994su} and references therein.\\[-10pt]}
Indeed, the line element for such curves reduces to:
\be
ds^{2}_{d+2}\Big|_{z,x^{i}=\, const.}\ =\ -d\beta(t)^{2} + {\beta(t)^{2} \over z^{2}} \( dz^{2} + g_{ij}dx^{i}dx^{j} \) \Big|_{z,x^{i}=\, const.}\ =\ -d\beta(t)^{2}\ .
\ee
For later use in the holographic renormalization of the action, it is also useful to rewrite such relation in the form:
\be\label{beta}
1 = \dot{\beta}/N = n^{\mu}\partial_{\mu}\beta = -\partial^{\mu}\beta\,\partial_{\mu}\beta\ .
\ee
These Eulerian observers can in turn be defined as those for whom our constant time hypersurfaces represent locally the set of events that are simultaneous.\footnote{
See also section 3.3 of \cite{Gourgoulhon:2007} for further details.\\[-10pt]}
In our case, such Eulerian worldlines are in fact geodesics with affine parameter $\beta$. Indeed, the acceleration of such worldlines is given by \cite{Gourgoulhon:2007}:
\be
a^{\mu} = n\cdot\na n^{\mu} = h^{\mu\nu}\partial_{\nu}\log N(t) = 0\ ,
\ee
where $h^{\mu\nu} = G^{\mu\nu} + n^{\mu}n^{\nu}$ is the induced metric of the surfaces of constant $t$ and hence its contraction with the gradient of the lapse vanishes since the latter is a pure function of the time coordinate. In the case of Minkowski space, such geodesics coincide with those found in \eqref{mink_geod} defining inertial particles in flat space. The scalar $\beta(t)$ therefore measures the invariant distance between points on different time slices connected by geodesics orthogonal to the slices.\\

\section{Holographic reconstruction of spacetime}\label{renormalization}

\subsection{Outline}\label{prescription}

\qquad In the previous section we obtained our spacetime asymptotics in the form \eqref{metric} with the expansion \eqref{FG_2}. In other words, after fixing the gauge freedom associated to a choice of coordinates, we found asymptotically the time evolution of our class of metrics. By construction, our bulk solution is foliated by conformally compact, asymptotically Einstein hypersurfaces of negative scalar curvature with a conformal boundary at $\del \mH$. Each such surface admits a dual description in terms of a $d$--dimensional Euclidean conformal field theory at $\del \mH$ and we would like to identify in this family of field theories the data necessary to the reconstruction of the bulk metric \eqref{metric}. In the previous section we found that, after choosing our time coordinate represented by a choice of $N(t)$, our spacetime asymptotics in the neighbourhood of the initial Cauchy surface is determined by the conformal structure of $\del \mH$ (the conformal class $[g_{(0)}]$) up to order $z^{d}$, excluding.  Moreover, we found that it is possible to move past such order and reconstruct the bulk metric near the surface up to very high order\footnote{
The obstacle to the reconstruction up to all orders is associated to the Fefferman-Graham coordinates \eqref{FG} on the surface. Once we have gauge-fixed our coordinate system on the surface, the gauge-fixing condition is valid only in a thickening near the boundary of the surface. This issue is analogous to the Gribov ambiguity in gauge theories and the absence of global gauge conditions, where gauge choices only hold in a neighbourhood of a gauge orbit.}
from the knowledge of the normalizable mode $g_{(d)}$. On each time slice, this mode is associated to the holographic energy tensor of the respective dual field theory on the conformal boundary \cite{deHaro:2000xn}. As we now briefly review for convenience, this result follows from the standard AdS/CFT prescription \cite{Witten:1998qj,Gubser:1998bc}, which identifies the supergravity partition function in asymptotically locally (Euclidean) AdS spaces with the generating functional of QFT correlation functions. The former is a functional of the boundary configurations $\phi_{(0)}$ of bulk fields and these are identified as sources for gauge-invariant operators $\op$ in the dual field theory. For a weakly coupled gravitational theory, one can work in a saddle-point approximation and take the gravitational on-shell action as the generating functional $W$ of connected QFT correlation functions:
\be\label{duality}
W[\phi_{(0)}]\ =\ \log\ \< \exp \( -\int_{\del \mH} \phi_{(0)}\op \) \>_{QFT}\ =\ \log\, Z_{{\rm SUGRA}}[\phi_{(0)}]\ \sim\, -S^{{\rm onshell}}[\phi_{(0)}]\ ,
\ee
where $S$ is the gravitational action in an asymptotically (E)AdS space and $\del \mH$ its conformal boundary. One can then obtain in particular the expectation value and correlation functions of the QFT energy tensor by functionally differentiating $S^{{\rm onshell}}$ with respect to the induced metric at $\del \mH$ representing the boundary configuration of the asymptotically AdS metric tensor. From the Hamilton-Jacobi theory, it follows that the on-shell action is a Hamilton principal functional and therefore that its first variation with respect to the induced metric results in the canonical momentum of the boundary, also known as the Brown-York quasi-local energy tensor \cite{Brown:1992br}. This tensor, however, in its bare form is ill-defined since both sides of \eqref{duality} suffer from divergences. One then proceeds by introducing a regulating boundary in the asymptotically AdS space and renormalizing the gravitational action with a set of covariant counterterms \cite{deHaro:2000xn}. Finally, since the conformal boundary is a region of the conformal embedding, one computes the Brown-York tensor of the regulating surface in the embedding from the renormalized action and in the end removes the regulator by taking the limit as the surface tends to the boundary. The renormalized Brown-York tensor obtained in this way corresponds by construction to the vacuum expectation value of the dual field theory energy tensor as computed at strong coupling from the renormalized generating functional $W^{ren}$ of the QFT. An explicit computation then shows that such one-point function indeed coincides exactly with the normalizable mode $g_{(d)}$ for odd $d$ and is equal to $g_{(d)}$ plus local functionals of the source $g_{(0)}$ for even values of $d$. See \cite{deHaro:2000xn} for further details.\\

The data necessary to the reconstruction of the bulk metric therefore consists of the modes $g_{(0)}$ (or any representative of the conformal structure $[g_{(0)}]$ at $\del \mH$) and $g_{(d)}$. While the former should be identified as the source for each conformal field theory stress tensor, the latter should be mapped to the expectation values of the stress tensors and in this section we will consider replacing the action on the right hand side of \eqref{duality} by the gravitational action for our class of Ricci-flat spacetimes in an attempt to confirm these entries in the holographic dictionary. By following the steps just outlined above, we will show that it is possible to reproduce the expectation values and Ward identities of the stress tensors of the field theories that reside at $\del \mH$ and to identify in this family the data necessary to the reconstruction of our spacetime asymptotics. Such results seem to support a non-trivial extension of the prescription \eqref{duality} to Ricci-flat embeddings of asymptotically hyperbolic manifolds. We will begin with the holographic renormalization of the gravitational on-shell action and then deduce the renormalized Brown-York tensor at $\del \mH$ which should correspond, for each fixed value of $t$, to the expectation value of the stress tensor of each field theory.

\subsection{Renormalization}

\qquad In order to renormalize the gravitational action, we begin by considering our spacetime as the region bounded by a (regulating) timelike hypersurface $\{ z=\e \}$. Such regulating boundary corresponds to one of the surfaces of constant $z$ in figure \ref{last_penrose} and in the end we will take the limit $\e \to 0$. In our coordinate system \eqref{metric}, we approach $\del \mH$ under such limit for finite, non-zero values of $\beta(t)$ (recall that $(\Im^{+},\mH) = \{z=(0,0)\, , \beta(t)=(+\infty,0)\}$ and $\beta(t) \in\, ]0,+\infty[$ ). The gravitational action is then given by:
\be\label{EH_S}
2\kappa^{2}_{d+2}\,S\ =\ \int_{\mM} d^{d+2}x\,\sqrt{G}\, R[G] + 2 \int_{z = \e} d^{d+1}x \sqrt{q}\, Q^{A}_{A}\ ,
\ee
where: $2\kappa^{2}_{d+2} = 16\pi G_{N}$ and where $q_{AB}$ and $Q_{AB}$ are the induced metric and the extrinsic curvature on the regulating boundary with coordinates $x^{A} = (t,x^{i})$. Given a spacetime with a timelike boundary $\{ z = \epsilon\}$, the Brown-York tensor, or the canonical momentum of the boundary, is obtained from the action as follows. One begins with a canonical decomposition of the spacetime metric $G_{\mu\nu}$:
\be
ds^{2}_{d+2}\ =\ G_{\mu\nu}dx^{\mu}dx^{\nu}\ =\ M^{2}dz^{2} + q_{AB} \( dx^{A} + U^{A}dz \) \( dx^{B} + U^{B} dz \)\ .
\ee
In our particular case we would have: $M = \beta(t)/z$ and $U^{A} = 0$. The extrinsic curvature $Q_{AB} = (2M)^{-1} \( \partial_{z} - \Lie_{U} \) q_{AB}$ and its extension to the spacetime is given in the same fashion as in \eqref{extension}. One then proceeds by rewriting the action in the canonical form using the Gauss-Codazzi identities (see \eg \cite{wald}):
\begin{align}
2\kappa^{2}_{d+2}\, S &= \int_{\mM} d^{d+2}x\, M\sqrt{q} \bigg( R[q] + Q^{2} - Q\cdot Q - 2 \na_{\mu} \( m^{\mu}Q - a^{\mu} \) \bigg) 
+ 2 \int\limits_{z=\e} d^{d+1}x \sqrt{q}\,Q\ \nn\\
&= \int_{\mM} d^{d+2}x\, M\sqrt{q} \bigg( R[q] + Q^{2} - Q\cdot Q \bigg)\ =\ 2\kappa_{d+2}^{2} \int dz\, L[q,q',M,U]\ , \label{canonical_S}
\end{align}
where the acceleration vector $a^{\mu} = m\cdot\na m^{\mu}$ and where $m = Mdz$ is the unit normal to the surfaces of constant $z$. Note that $a^{\mu}m_{\mu}=0$. Also, $q':=\partial_{z}q$ and $L$ is the canonical Lagrangian. We emphasize that it is the on-shell action in the canonical form that is a Hamilton principal functional and therefore a functional of the boundary configuration of the spacetime metric. Furthermore:
\be
\delta G_{\mu\nu}\ =\ \Big(2\, m_{\mu} m_{\nu}/M \Big) \delta M + \Big(2\, m_{(\mu}q_{\nu)A}/M \Big) \delta U^{A} + \Big( q_{\mu}^{A}q_{\nu}^{B} \Big)\, \delta q_{AB}\ ,
\ee
and hence the action $S = S[G]$ is a functional of the canonical fields $M,U^{A}$ and $q_{AB}$. The variation of $S$ is then given by:
\begin{align}
\delta S\ &= \int dz \int d^{d+1}x \( {\delta L \over \delta q_{AB}}\, \delta q_{AB} + {\delta L \over \delta q_{AB}'}\, \delta q_{AB}' + ... \) \nn\\[5pt]
&= \int dz \int d^{d+1}x \( {\delta L \over \delta q_{AB}} - {d \over dz}{\delta L \over \delta q_{AB}'} \) \delta q_{AB} + ... + \int\limits_{z = \e} d^{d+1}x {\delta L \over \delta q_{AB}'} \delta q_{AB}\ , \label{var_S}
\end{align}
where the ellipses denote the variation of the Lagrangian with respect to the lapse and shift $M$ and $U^{A}$ which are Lagrange multipliers in the canonical formalism. Since \makebox{$\delta L / \delta q_{AB}'(z=\e)$} is by definition the canonical momentum $\pi^{AB}$ of the regulating boundary that is conjugate to $q_{AB}$, and since the remaining terms in \eqref{var_S} represent the equations of motion and the Hamiltonian and diffeomorphism constraints, we find:
\be
\delta S \Big|_{{\rm on-shell}}\ =\ \int_{z = \e} d^{d+1}x\, \pi^{AB}\delta q_{AB}\ .
\ee
A quick computation using \eqref{canonical_S} shows that $\pi^{AB} = - \sqrt{q} \( Q^{AB} - q^{AB}Q \)/2\kappa_{d+2}^{2}\ $ and hence:
\be
{2\kappa_{d+2}^{2} \over \sqrt{q}}\, {\delta S^{{\rm onshell}} \over \delta q^{AB}(z=\e)}\ =\ Q_{AB} - q_{AB}Q\ ,
\ee
which represents the Brown-York tensor of the regulating boundary. An explicit computation of this tensor using our solution \eqref{metric} reveals that it diverges under the limit $\e \to 0$ and therefore we need to renormalize the action \eqref{canonical_S}.\footnote{
It should be emphasized that it is the canonical action \eqref{canonical_S}, as opposed to \eqref{EH_S}, that is renormalized, since it is the former that is the functional of the boundary configuration of the metric. If the lapse $M=M(z)$, then the acceleration $a^{\mu}$ is identically zero and therefore both forms of the action exhibit the same divergences as evaluated by an examination of the asymptotic form of the integrands. In such case, it is irrelevant which of the two actions is renormalized. Indeed, the gauge choice $M=M(z)$ is the standard gauge in conventional AdS/CFT and hence one can renormalize directly the gravitational action without any canonical decomposition. For a different gauge choice, a canonical decomposition of the action is required (or then, the addition of a total derivative $-2\na_{\mu}a^{\mu}$ to the undecomposed action). In the language of AdS/CFT, the UV divergences of the QFT partition function are mapped to the IR divergences of the on-shell gravitational action in the canonical form. In our case, from the spacetime asymptotics \eqref{metric} we find that $M=M(z,t)$ and hence $a^{\mu} = -q^{\mu\nu}\partial_{\nu}\log M = n^{\mu}/\beta \neq 0$. See also \cite{Papadimitriou:2004ap,Papadimitriou:2010as}.\\[-5pt]
}
In order to do so, we use our solution \eqref{metric} and the asymptotic series \eqref{FG_2} and begin by evaluating the integral on-shell\footnote{ 
The integration limits in the $z$-integral are as follows: $\int^{\e}_{z_{0}} dz$, for some constant $z_{0}>\e$.
} 
and reading the terms that diverge as $\e \to 0$. We find that the divergent terms are of the form:
\be\label{div_terms}
\int_{z=\e} d^{d+1}x \( \sum\limits_{n>0}^{d} A_{n}\,\e^{-n} + \mA \log \e \)\ ,
\ee
where the coefficients $A_{n}$ and $\mA$ are local functionals of $g_{(0)ij}$ and $\beta$. We then invert the expansion \eqref{FG_2} in order to express $g_{(0)ij}$ order by order in $\e$ in terms of $g_{ij}$ and therefore covariantly in terms of $\g_{ij} := \beta(t)^{2}/\e^{2} g_{ij}$, where $\g$ is the induced metric on the surfaces \makebox{$\{z=\e\, ,\, t=constant \}$}, and replace the inverted expansion $g_{(0)} = g_{(0)}[\g,\e,\beta]$ in the functionals $A_{n}$ and $\mA$.  The next step is to introduce the projector $\g_{AB} := q_{AB} + n_{A}n_{B}$ onto the surfaces of constant time with unit normal $n_{A}$ and use the standard identities from the theory of embedded hypersurfaces to extend the scalar functionals of $\g_{ij}$ to scalar functionals of $\g_{AB}$ on the submanifold $\{ z=\e\}$. The divergent terms written in this way can then be minimally subtracted from the action by introducing a preliminary counterterm integral at ${z = \e}$ consisting of minus such divergent terms. These counterterms are given by:
\begin{align}
2\kappa_{d+2}^{2}\,S^{pre-CT} &= \int_{z=\e} d^{d+1}x \sqrt{q} \( 2(d-1) \beta^{-1} + {\beta \over d-2}\, R^{A}_{A}[\g] \right.  \nn\\[5pt] 
&\left.  +\, {\beta^{3} \over (d-4)(d-2)^{2}} \( R_{AB}[\g]R^{AB}[\g] - {d \over 4(d-1)} \(R^{A}_{A}[\g]\)^{2} \) + ... + \beta^{d-1}\mA_{(d)} \log \e \) \nn\\[10pt]
&= \int_{z=\e} d^{d+1}x \sqrt{q}\, \bigg( \sum\limits_{n=0}^{[d/2-1]} C_{n}\, \beta^{2n-1} + \beta^{d-1}\mA_{(d)} \log \e \bigg)\ , \label{pre_CT}
\end{align}
where we have written explicitly the counterterms $C_{n}$ up to $d = 6$. The notation $[d/2-1]$ represents the integer value of $d/2-1$ rounded up. The coefficient $\mA_{(d)}$ will be proportional to the conformal anomalies of the field theories, it vanishes for odd $d$ and for even $d$ up to $d=4$ is given by:
\be
\mA_{(2)} = -R^{A}_{A}[\g]\quad ,\quad \mA_{(4)} = -{1 \over 8} \( R_{AB}[\g]R^{AB}[\g] - {1 \over 3} \(R^{A}_{A}[\g]\)^{2} \) \ .
\ee
Notice that, on the surfaces of constant $t$, these preliminary counterterms agree with those found in \cite{deHaro:2000xn} in the context of AdS/CFT. Furthermore, and as in AdS holographic renormalization, these counterterms break invariance of the action with respect to diffeomorphisms involving the radial coordinate $z$ due to the explicit dependence on the regulator $\e$ and this will be associated as usual to the conformal anomalies of the field theories.

Due to the dependence of $S^{pre-CT}$ on $\beta(t)$, this counterterm action is not yet fully covariant. We can then covariantise it by introducing a boundary Lagrange multiplier $\lambda$ for $\beta$. The simplest choice is to use the identity \eqref{beta} and add to \eqref{pre_CT} the Lagrange multiplier:\footnote{
The action \eqref{lagr_mult} for $\beta$ is identical to the action for a pressureless perfect fluid, or dust, at the boundary, where $\l$ is proportional to the fluid's rest mass density and $\beta$ plays the role of the fluid's propertime \cite{Brown:1994py}. This feature simply follows from the fact that the set of Eulerian worldlines described at the end of section \ref{regionII} for which $\beta$ is the propertime behaves as a congruence of dust particles. Equation \eqref{EOM_lambda} is a continuity equation for the rest mass current and, in the absence of the counterterm action \eqref{pre_CT}, it expresses the conservation of the total mass.}
\be\label{lagr_mult}
2\kappa_{d+2}^{2}\, S^{\l}\ =\ \int_{z = \e} d^{d+1}x \sqrt{q}\, \lambda \( 1 + q^{AB}\partial_{A}\beta\partial_{B} \beta \)\ .
\ee
We then treat $\beta$ as a dynamical field at the boundary with respect to which the variation of the renormalized action $S^{ren} := S + S^{pre-CT} +  S^{\l}$ should vanish. Such variation then results in the configuration for $\l$:
\bea
{\delta S^{ren} \over \delta \l} = 0\quad &\Lra&\quad q^{AB}\partial_{A}\beta\partial_{B}\beta\ =\ -1\ , \label{EOM_beta}\\[5pt]
{\delta S^{ren} \over \delta \beta} = 0\quad &\Lra&\quad \partial_{A} \( \sqrt{q}\, \l\, \partial^{A}\beta \) =\ \hf\, \sqrt{q} \(  \sum\limits_{n=0}^{[d/2-1]} (2n-1)\,C_{n}\, \beta^{2(n-1)} \right. \nn \\
&& \hspace{1.7in} +\, (d-1)\beta^{d-2}\mA_{(d)} \log \e \Bigg)\ . \label{EOM_lambda}
\eea
By using our solution \eqref{metric}, now written as:
\be
ds^{2}_{d+2} = - N(t)^{2}dt^{2} + {\bar{\beta}(t)^{2} \over z^{2}} \( dz^{2} + g_{ij}dx^{i}dx^{j} \)\quad :\quad \bar{\beta}(t) := \int dt N(t)\ ,
\ee
and looking for solutions $\beta = \beta(t)$, we obtain:
\bea
\beta &=& \bar{\beta}(t)\ ,\\[5pt]
2\,\l &=& -\sum\limits_{n=0}^{[d/2-1]} {2n-1 \over d-1}\,C_{n}\,\bar{\beta}(t)^{2n-1} - \bar{\beta}(t)^{d-1} \mA_{(d)} \log \e \ , \label{sol_lambda}
\eea
where we have set the integration constants to zero. The final renormalized action is therefore given by:
\bea\label{S_ren}
2\kappa^{2}_{d+2}\, S^{ren} &=& \int\limits_{\mM} d^{d+2}x\, M\sqrt{q}\, \bigg( R[q] + Q^{2} - Q\cdot Q \bigg) \nn\\[5pt]
&+& \int\limits_{z=\e} d^{d+1}x \sqrt{q}\, \bigg( \sum\limits_{n=0}^{[d/2-1]} C_{n}\, \beta^{2n-1} + \beta^{d-1}\mA_{(d)} \log \e \bigg) \nn \\[5pt]
&+& \int\limits_{z = \e} d^{d+1}x \sqrt{q}\, \lambda \( 1 + |\partial\beta|^{2}\)\ .
\eea
The on-shell value of $\delta S^{ren}$ is then given by:
\bea
2\kappa_{d+2}^{2}\, \delta S^{ren}\Big|_{{\rm on-shell}}&=& \int\limits_{z=\e} d^{d+1}x\, \sqrt{q}\, \Big( Q_{AB} - q_{AB} Q \Big) \delta q^{AB} \nn \\[5pt]
&-& \hf\, \int\limits_{z=\e} d^{d+1}x \sqrt{q}\, \bigg( \sum\limits_{n=0}^{[d/2-1]} C_{n}\, \bar{\beta}^{2n-1} + \bar{\beta}^{d-1}\mA_{(d)} \log \e \bigg) q_{AB}\, \delta q^{AB} \nn \\[5pt]
&+& \int\limits_{z=\e} d^{d+1}x \sqrt{q}\, \bigg( \sum\limits_{n=0}^{[d/2-1]} \bar{\beta}^{2n-1}\, {\partial C_{n} \over \partial q^{AB}} + \bar{\beta}^{d-1}\, \log \e\, {\partial \mA_{(d)} \over \partial q^{AB}} \bigg) \delta q^{AB} \nn \\[5pt]
&+& \int\limits_{z = \e} d^{d+1}x \sqrt{q}\, \lambda \( \partial_{A} \bar{\beta} \partial_{B} \bar{\beta}\) \delta q^{AB}\ . \label{var_S_ren}
\eea
Using equation \eqref{sol_lambda} for $\lambda$ and the identity: $\partial_{A}\bar{\beta} = - n_{A}\,$, the last integral in \eqref{var_S_ren} is given by:
\begin{align}
\int\limits_{z = \e} d^{d+1}x \sqrt{q}\, \lambda \( \partial_{A} \bar{\beta} \partial_{B} \bar{\beta}\) \delta q^{AB}\ =\ -\hf \int\limits_{z=\e} d^{d+1}x \sqrt{q}\, &\Bigg( \sum\limits_{n=0}^{[d/2-1]} {2n-1 \over d-1}\,C_{n}\,\bar{\beta}^{2n-1} \nn\\
&\quad  +\, \bar{\beta}^{d-1} \mA_{(d)} \log \e \Bigg) n_{A}n_{B}\, \delta q^{AB}\ .
\end{align}
In this way, the renormalized Brown-York tensor is given by (we drop the bar notation over $\beta$ from now on):
\bea\label{ren_BY}
{2\kappa_{d+2}^{2} \over \sqrt{q}}\, {\delta S^{ren}_{onshell} \over \delta q^{AB}(z=\e)} &=& Q_{AB} - q_{AB} Q + \sum\limits_{n=0}^{[d/2-1]} \beta^{2n-1} \( {\partial C_{n} \over \partial q^{AB}} - \hf\, C_{n} \( q_{AB} + {2n-1 \over d-1}\, n_{A}n_{B} \) \) \nn\\[5pt]
&+& \beta^{d-1} \( {\partial \mA_{(d)} \over \partial q^{AB}} - \hf\, \mA_{(d)} \g_{AB} \) \log \e\ .
\eea
As an exercise, for $d+2=4$ we obtain:
\be
{2\kappa_{4}^{2} \over \sqrt{q}}\, {\delta S^{ren}_{onshell} \over \delta q^{AB}(z=\e)}\ =\ Q_{AB} - q_{AB} Q - 2\beta^{-1}q_{AB} + \beta^{-1}\g_{AB}\ ,
\ee
where the term $(R_{AB}[\g] - \hf\g_{AB}R[\g])\log\e\, $ vanishes identically for $d=2$.
\\

\subsection{Holographic stress tensors}\label{holog_energy-tensor}

\noindent Following \cite{Brown:1992br}, we now decompose the Brown-York tensor into the spatial stress tensor $s_{ij}$ and the momentum and energy densities $j_{i}$ and $\ve$:
\bea
&&s_{ij}\ :=\ {2 \over N\sqrt{\g}} {\delta S^{ren}_{onshell} \over \delta \g^{ij}}\ =\ \g_{i}^{A}\g_{j}^{B} \( {2 \over \sqrt{q}} {\delta S^{ren}_{onshell} \over \delta q^{AB}} \)\ , \label{s_ij}\\[5pt]
&&j_{i}\ :=\ {1 \over \sqrt{\g}} {\delta S^{ren}_{onshell} \over \delta V^{i}}\ =\ \g_{i}^{A}n^{B} \( {2 \over \sqrt{q}} {\delta S^{ren}_{onshell} \over \delta q^{AB}} \)\ , \label{j_i}\\[5pt]
&&\ve\ :=\ - {1 \over \sqrt{\g}} {\delta S^{ren}_{onshell} \over \delta N}\ =\ -n^{A}n^{B} \( {2 \over \sqrt{q}} {\delta S^{ren}_{onshell} \over \delta q^{AB}} \)\ , \label{vareps}
\eea
where: $q_{AB}dx^{A}dx^{B} = - N^{2}dt^{2} + \g_{ij} \( dx^{i} + V^{i}dt \) \( dx^{j} + V^{j}dt \)$ and: $\sqrt{q} = N \sqrt{\g}$. For our spacetime solution, the non-trivial components are the spatial stress and energy density; the contraction of the Brown-York tensor \eqref{ren_BY} with $\g_{i}^{A}n^{B}$ vanishes identically, resulting in a vanishing momentum $j_{i}$.

The expectation value of the stress tensor of each field theory is now obtained by computing the spatial components $s_{ij}$ of the renormalized Brown-York tensor in the conformal embedding and taking the limit as the regulating surface $\{z = \e\}$ tends to the boundary $\{ z = 0\}$. Recall from \eqref{conf_comp} that our defining function $\rho = z/\beta$. By factorising the latter, our spacetime solution reads as:
\be
ds^{2}_{d+2} = {\beta(t)^{2} \over z^{2}} \( dz^{2} - z^{2}N_{(0)}^{2}dt^{2} + (g_{(0)ij} + \op(z^{2})\, )\, dx^{i}dx^{j} \)\ ,
\ee
with $N_{(0)} := N(t) / \beta(t)$. The metric $\tilde{G}$ of the conformal embedding is given by $\tilde{G}_{\mu\nu} = (z/\beta)^{2}G_{\mu\nu}$ and the expectation value of the operator dual to $g_{(0)}$ is therefore obtained as:\footnote{
Here and in equation \eqref{def_E} we have performed the intermedium step: $\delta S/\delta \tilde{q}^{AB} = (z/\beta)^{2} \delta S / \delta q^{AB}$, where: $\tilde{q}_{AB}:= (z/\beta)^{2}q_{AB}$, and performed the decomposition of $\delta/\delta \tilde{q}^{AB}$ in the same way as in \eqref{s_ij}--\eqref{vareps}.\\[-5pt]
}
\be
\< T_{ij} \>\ =\ {2 \over N_{(0)}\sqrt{g_{(0)}}} {\delta S^{ren}_{onshell} \over \delta g_{(0)}^{ij}}\ =\  \lim_{\e \to 0} \( \beta \(\beta / \e\)^{d-2} s_{ij} \)\ . \label{def_T} \\[5pt]
\ee
From the perspective of the standard AdS/CFT dictionary, we cannot interpret $N_{(0)}$ in the usual sense as a source term for some dual operator in each field theory (note that $N_{(0)}$ cannot be switched off). We cannot interpret it as a source for the time-component of the energy tensor at $z=0$ either, due to the different asymptotic behaviours of the spatial and time components of the metric. Nevertheless, we can still define the renormalized quantity:
\be
\mE\ :=\ - {1 \over \sqrt{g_{(0)}}} {\delta S^{ren}_{onshell} \over \delta N_{(0)}}\ =\ \lim_{\e \to 0} \( \beta \( \beta / \e \)^{d} \ve \)\ . \label{def_E}
\ee
An explicit computation of $\<T_{ij}\>$ and $\mE$ using our spacetime solution results in the following expectation values:\footnote{
Notice that the term $Q_{AB}$ as well as: $\partial C_{n} / \partial q^{AB}$ and: $(\partial \mA_{(d)} / \partial q^{AB} - \hf \mA_{(d)} \g_{AB}) \log \e$ in \eqref{ren_BY} are purely spatial and therefore vanish when contracted with $n^{A}$. The expression for $\mE$ follows from an explicit computation of \eqref{def_E} for each value of $d$ and has been verified up to $d=6$.\\[-5pt]}
\bea
\< T_{ij} \> &=& {d\, \beta(t)^{d} \over 2\kappa^{2}_{d+2}} \( g_{(d)ij} + X_{ij}[g_{(0)}] \)\ , \label{vev_T} \\[5pt]
\mE  &=& {1 \over d-1}\, \< T^{i}_{i} \>\ , \label{vev_E}
\eea
where $X_{ij}$ is a functional of $g_{(0)}$, it vanishes for odd values of $d$ and its explicit expression for even values depends on $d$. Also, the trace of $\<T_{ij}\>$ is taken with respect to $g_{(0)}$. Equation \eqref{vev_T} represents the expectation value of the spatial stress tensor of each field theory described holographically by a hypersurface of constant $t$ and it coincides with the holographic stress tensor found in \cite{deHaro:2000xn} in the context of AdS/CFT. Equation \eqref{vev_E} identifies $\mE$ with the Weyl anomaly of each field theory.\footnote{
As an observation, by using the definitions \eqref{def_T} and \eqref{def_E}, equation \eqref{vev_E} can be rewritten as:
\be
\( -2\,g_{(0)}^{ij}\, {\delta \over \delta g_{(0)}^{ij}} + (1-d)\, N_{(0)}\, {\delta \over \delta N_{(0)}} \) S^{ren}_{onshell}\ =\ 0 , 
\ee
and which represents the holographic Callan-Symanzik equation for a anomaly-free CFT deformed by a source $N_{(0)}$ for a relevant scalar operator of dimension one. The Weyl anomaly of each field theory can therefore also be seen as that created by a non-vanishing vacuum expectation value of such an operator. Notice that, in such case and by using \eqref{diffeo_ward}, the diffeomorphism Ward identity would also be satisfied since $\partial_{i}N_{(0)} = 0$ when on-shell.\\[-5pt]
}
The holographic Ward identities for the field theories are given by:
\bea
&&\cov_{j}\, \< T^{j}_{\ i} \>\ =\ 0\quad \forall d , \label{diffeo_ward}\\[5pt]
&&\< T^{i}_{i} \>\ =\ 0\quad :\ d = 2n+1\ ,\ n \in \Integer\ , \label{trace_ward_odd}
\eea
where $\cov_{j}$ is the covariant derivative associated to $g_{(0)}$. For even values of $d$, the trace of the stress tensor depends on $d$. For $d=2$, we find:\footnote{
If $d+2=4$, then our (asymptotic) solution \eqref{metric} represents the Lorentzian cone of a three-dimensional Einstein Riemannian manifold of negative scalar curvature. The latter is therefore the hyperbolic 3-space, up to possible global identifications, and hence \eqref{metric} is diffeomorphic to Minkowski spacetime.\\[-5pt]
}
\be
\< T_{ij} \>\ =\ {\beta(t)^{2} \over \kappa^{2}_{4}} \( g_{(2)ij} - g_{(0)ij} \Tr[g_{(0)}^{-1}g_{(2)}] \)\ .
\ee
The holographic Weyl anomaly in this case is given by:
\be
\< T^{i}_{i} \>\ =\ {\beta(t)^{2} \over 2\kappa^{2}_{4}}\, R[g_{(0)}]\ =\ {c \over 24\pi}\, R[g_{(0)}]\ ,
\ee
where the central charge of each 2-dimensional CFT is related to $\beta(t)$ as:
\be\label{beta_c}
\beta(t)\ =\ \ell_{P}' \sqrt{{c \over 6}}\qquad\ :\quad \ell_{P}'\, :=\, 2\, \sqrt{{\hbar\, G_{N} \over c_{0}^{3}}}\ ,
\ee
where we have reinserted\footnote{
Recall that $\hbar$ arises from the partition function \eqref{duality}.\\[-10pt]
}
the factors of $\hbar$ and $c_{0}$ and defined $\ell_{P}'$ as twice the usual convention for the Planck length in four bulk dimensions. As an exercise, for four dimensional Minkowski space \eqref{mink_II} in the cone coordinate (also known as Milne's space):
\be
ds^{2} = -dt^{2} + t^{2} \Bigg( {dz^{2} \over z^{2}}+ {1 \over z^{2}} \({1-z^{2} \over 2}\)^{2} d\Omega^{2} \Bigg)\ ,
\ee
we obtain:
\bea
&&\< T_{ij} \>\ =\ {t^{2} \over 2\kappa^{2}_{4}} \( 4g_{(0)ij}\)\ ,\\[5pt]
&&c / 6\ =\ \(t / \ell_{P}'\)^{2}\ ,
\eea
with $4g_{(0)ij}dx^{i}dx^{j} = d\Omega^{2}$ the metric on the $S^{2}$.\\

Equation \eqref{beta_c} associates the spectrum of central charges to the bulk time coordinate essentially in a gauge-invariant way, \ie it does not depend on a particular choice of time coordinate due to the invariant meaning of $\beta(t)$, and it requires, either a notion of locality in the spectrum, or a discretisation of time distances in units of the Planck length. More generally, in even dimensions, the time coordinate is mapped to the coefficients, or central charges, of the Euler density and Weyl invariant in the conformal anomalies, also known as type A and B anomalies \cite{Deser:1993yx}.\footnote{
Recall that, holographically and in the absence of higher curvature corrections, these coefficients are related, see \eg \cite{Henningson:1998gx,Henningson:1998ey}
} 
In AdS/CFT, the renormalization group equations are local in the energy scale and this property is consistent with the notion of (coarse) locality in the bulk radial direction, hence we obtain a precise matching on both sides of the duality. In our case, on the other hand, the interrelationship between the different field theories in the family, and in particular between their central charges, is not clear. Consistency of our framework therefore requires a notion of locality in the spectrum so that locality in the time direction is recovered. In order to understand more comprehensively the correlation between the field theories required by the duality, as well as the role of the time coordinate in the family, it seems necessary to study the full group of diffeomorphisms in the bulk that preserves the form of the metric \eqref{metric}. Such diffeomorphisms contain a subgroup involving the time coordinate \cite{deBoer:2003vf} which should act at $\del \mH$ as a particular type of transformation between the different field theories. By construction, this family should be invariant under such transformations and an identification of such symmetry group should allow a better understanding of the way the field theories are connected.\\

Before ending this section and as an exercise, we can consider a simple example of a family of two-dimensional CFTs with a continuous parametrisation of the central charge and which we will take to be a family of Liouville field theories. Classical Liouville theory is the theory of two-dimensional conformal manifolds, or Riemann surfaces. Suppose we have a two-dimensional Riemannian manifold with metric:
\be\label{tilde_g}
ds^{2}\ =\ \tilde{g} _{ij} dx^{i}dx^{j}\ :=\ e^{2b\phi} g_{(0)ij}dx^{i}dx^{j}\ ,
\ee
where $\tilde{g}$ is a representative of the conformal structure $[\tilde{g}]$ of some Riemann surface, $b$ is a dimensionless constant and $g_{(0)}$ is an arbitrary ``background" metric. The field $\phi(x)$ is known as the Liouville field. The scalar curvature of $\tilde{g}$ and $g_{(0)}$ are then related as: $R[\tilde{g}] = e^{-2b\phi} \( R[g_{(0)}] - 2b \square \phi \)$, where $\square$ is the Laplacian with respect to $g_{(0)}$. Since the Riemann surface is endowed with a conformal structure, we can take the representative $\tilde{g}$ to be a metric of constant curvature and write: $R[\tilde{g}] = -8 \pi \mu b^{2}\ :\ \mu > 0$. In this way, we find: $\square \phi = 4\pi\mu b e^{2b\phi} + (2b)^{-1} R[g_{(0)}]$. The action for this theory is the classical Liouville action  \cite{Zamolodchikov:1995aa,Zamolodchikov:2001ah}:
\be
S\ =\ {1 \over 4\pi} \int d^{2}x \sqrt{g_{(0)}} \( g_{(0)}^{ij}\partial_{i}\phi\partial_{j}\phi + 4\pi\mu e^{2b\phi} + Q R[g_{(0)}] \phi \)\ ,
\ee
where the so-called background charge $Q = b^{-1}$. The continuous parameter $b$ is either real of purely imaginary, corresponding respectively to a metric $\tilde{g}$ of negative or positive curvature.\footnote{
In the latter case, the Liouville field is redefined as $\phi \to -i\phi$, resulting in a negative sign in the kinetic term in the action.
}
Since we are only interested in the conformal structure $[\tilde{g}]$, we require that $\phi$ transforms as: $\phi \to \phi - (Q/2) \log \Omega$ under the conformal transformation: $g_{(0)} \to \Omega\, g_{(0)}$, such that $\tilde{g} \to \tilde{g}$. The equation of motion for $\phi$ is therefore invariant under the combined transformations, which implies that the action is also invariant up to boundary terms, as long as $Q=b^{-1}$. Since every two-dimensional metric is conformally flat, we can now set \makebox{$g_{(0)} = \mathbbm{1}$} (which just corresponds to a redefinition of $\phi$ in \eqref{tilde_g}) and then change to complex coordinates: $\(z = x^{1}+ix^{2}\, ,\, \bar{z} = x^{1}-ix^{2}\)$ such that:
\be\label{complex}
ds^{2}\ =\ e^{2b\phi} dzd\bar{z}\ .
\ee
In the gauge \eqref{complex}, the non-vanishing components of the holomorphic Liouville energy tensor (obtained from $\delta S / \delta g_{(0)}$) are given by:
\bea
T(z) &=& - \( \partial \phi \)^{2} + Q \partial^{2}\phi\ ,\\
\bar{T}(\bar{z}) &=& - \( \bar{\partial} \phi \)^{2} + Q \bar{\partial}^{2}\phi\ ,
\eea
where we have used the equation of motion for $\phi$. Under the holomorphic transformation: $z \to \omega(z)\ ,\ \phi \to \phi - (Q/2) \log |\partial \omega|^{2}$, the energy tensor transforms as:
\be
T(z) \to \(\partial \omega\)^{-2} \( T(z) - {Q^{2} \over 2} \( {\partial^{3}\omega \over \partial \omega} - {3 \over 2} \( {\partial^{2} \omega \over \partial \omega} \)^{2}\) \)\ .
\ee
A comparison with the standard transformation law for $T(z)$ under a holomorphic conformal transformation yields the ``classical" central charge:
\be
c\ =\ 6\, Q^{2}\ .
\ee
This result can also be obtained by considering Liouville's theory on a cylinder and verifying that the Fourier components of the energy tensor satisfy the Virasoro Poisson bracket algebra with the above central charge (see \eg \cite{Seiberg:1990eb} with a slightly different notation). From the transformation law for the Liouville field, it follows that the fields $e^{2\alpha\phi}$ are primary with ``classical" dimension $\Delta = \alpha Q$, for some constant $\a$. When the Liouville theory is quantised, the background charge $Q$ receives corrections, as well as the central charge and the dimensions $\Delta$. After normal ordering the (off-shell) energy tensor, its Fourier components only satisfy the Virasoro commutator algebra (with the canonical commutation relations imposed on the Liouville field and momentum, see \cite{Curtright:1982gt}) if $Q = b^{-1} + b$. The Virasoro algebra then yields the central charge:
\be\label{Liouville_c}
c\ =\ 6\, Q^{2} + 1\ .
\ee
The classical limit corresponds to $b \to 0$. Finally, by considering the OPE of the operators $e^{2\alpha\phi}$ with the energy tensor, it follows that these are primary, now with dimension:
\be\label{delta_alpha}
\Delta = \alpha Q - \alpha^{2}\ .
\ee
In conventional AdS/CFT, a Liouville theory at the conformal boundary can be realised holographically by an asymptotically hyperbolic 3-space with a scalar field of mass $m^{2} = \a( Q - \a)\(\a( Q - \a)-2\)$ for each dynamical operator $e^{2\a \phi}$. Motivated by our results, we also expect that, by switching on some scalar field in our spacetime in four dimensions, we are able to capture holographically the dynamics of scalar operators in a family of two-dimensional CFTs at $\del \mH$. In \cite{deBoer:2003vf} it was brought to the attention that, in general, each (non-backreacting) scalar field in the bulk decomposes into an infinite set of massive scalars on each surface of constant time. This feature poses several problems to the holographic computation of correlators of scalar operators, in particular because we know that each massive scalar on a constant time slice is dual to a single scalar operator of definite scaling dimension in the dual field theory. Since we do not have a definite mass (or a unique scalar field) on a slice of constant time associated to a scalar field on the spacetime, we lose the correspondence between a bulk field and a single operator per field theory just as we had between the spacetime metric and the stress tensors. It seems possible, however, to bypass this issue and reproduce the correlators of one operator per field theory by setting up an initial value problem for the bulk scalar in a manner similar to the approach that we followed in the case of the spacetime metric. In this way, we restrict the solutions of the bulk wave equation to a subclass that suffices to capture the dynamics of the operators in the field theories.\footnote{
Of course that, in this way, we are not able to reconstruct holographically any solution of the bulk wave equation, but rather a subclass of such solutions. In the same way, we are not able to reconstruct holographically any asymptotically Ricci-flat metric, but rather the subclass of such metrics that admit an asymptotically hyperbolic hypersurface of constant mean curvature. It seems that such general classes of solutions contain more information than that that just a family of field theories in two dimensions less can provide. On the other hand, if we are only interested in reproducing the correlators of the operators of each field theory in the family, then the above approach should be sufficient.\\[-7pt]
}
Such particular solutions reduce on each slice to a single massive scalar and in this case, the mass of the field on a slice, and therefore the dimension of the dual operator, will be associated to the time-dependence of the bulk spacetime field, \ie to its derivatives with respect to $\beta(t)$. With a correspondence between a bulk solution and a single operator per field theory, we should be able to reproduce holographically the correlators of operators in the family of field theories. The holographic reconstruction of bulk matter fields in our class of Ricci-flat spacetimes will be addressed in future work \cite{Caldeira:WIP}.
\\

Returning to the case at hand, and in particular for a family of Liouville field theories at $\del \mH$, we have deduced that pure gravity in the bulk captures the dynamics of the stress tensors of each theory. By using equations \eqref{beta_c} and \eqref{Liouville_c}, it then follows that $\beta(t)$ is associated to the continuous background charge of the family as:
\be\label{beta_Q}
\beta(t)\ =\ \ell_{P}' \sqrt{Q^{2}+1/6}\ \sim\ \ell_{P}'\, Q\qquad (b \to 0)\ .
\ee
Equation \eqref{delta_alpha} then implies that the scaling dimensions of the primary operators of each field theory are also expressed in terms of $\beta(t)$. If a scalar field in the bulk is dual to a family of primary operators (each belonging to a different field theory) of the same scaling dimension $\Delta$, then different operators in this family must have different $\a$'s. From equation \eqref{delta_alpha} we find that: $\alpha(t) = Q/2 \pm \sqrt{(Q/2)^{2} - \Delta}\,$, where the time dependence of $\a$ is obtained from the relation \eqref{beta_Q}. The bulk scalar would therefore be dual to the family operator: $exp\Big[ \( Q(t) \pm \sqrt{Q(t)^{2} - 4\Delta} \) \phi_{t}\Big] $, where $\phi_{t}$ is the operator of the field theory with central charge $c(t)$. In the limit as $b \to 0$ (or at late times $\beta(t) \gg 1$), the family operator asymptotes to $exp\(2\beta\phi / \ell_{P}'\)$ or to the identity operator.\\

In the next section we will generalise our class of asymptotically Ricci-flat spacetimes, deduce their asymptotics and compute the modifications to the expectation values of the stress tensors. This is done for the following reason. From equation \eqref{vev_T} it follows that, up to a constant factor, different CFTs have the same expectation values of the stress tensors.\footnote{
The $n$-point correlators should also be the same. These are obtained by functionally differentiating $\< T_{ij} \>$ with respect to $g_{(0)}$ for an exact bulk solution and, since both the source and the normalisable mode $g_{(d)}$ are $t$-independent, such differentiation should not introduce any additional time dependence which would discriminate between different CFTs.}
This implies that the reconstruction of the spacetime asymptotics only requires the knowledge of the holographic stress tensor of a single CFT, besides the conformal structure $[g_{(0)}]$. In order to obtain different vevs for different field theories, a relative time-dependence is needed in the expressions \eqref{vev_T} for the expectation values which would discriminate between different CFTs. This could be achieved by requiring for example that $\partial_{t}g_{ij}\neq 0$ in the spacetime metric \eqref{metric}. Indeed, we can obtain different expectation values of the stress tensors for different field theories by considering subleading contributions to the asymptotic value of the initial data $K_{[0]}$. The solution to such Cauchy problem will be similar to our solution \eqref{metric}, but with a particular time-dependent $g_{ij}$. The reconstruction of the spacetime asymptotics in such case will then require the knowledge of a family of different holographic stress tensors. In the next section we will therefore allow an arbitrary initial extrinsic curvature $K_{[0]}$ away from the conformal boundary of the initial data surface such that it approaches $h_{[0]}$ only asymptotically.\\

\section{Corrections to the holographic stress tensors}\label{horizon}

\subsection{Generalisation of the initial data}

\qquad As in section \ref{regionII}, we start from an asymptotically hyperboloidal initial data set $( \Sigma, h_{[0]},K_{[0]} )$, with $(\Sigma,h_{[0]})$ of dimension $d+1$, and consider the first subleading orders of $K_{[0]}$ as we move away from the conformal boundary $\{z=0\}$ of the initial Cauchy surface. The asymptotic form of our initial data is therefore:
\bea
&&R_{ab}[h_{[0]}] = -{d \over \ell^{2}}\, h_{[0]ab}\quad , \label{data_1} \\[5pt]
&&K_{[0]ab} = {1 \over \ell}\,h_{[0]ab} + z^{-2+\a} \({d \over 2}\, \ell\, \mT_{ab}\) \quad :\quad \mT_{ab}(z,x) = \mT_{(0)ab}(x) + \op(z^{>0}) \quad , \label{data_2}
\eea
where we have now kept for convenience the constant curvature radius $\ell$ of $\Sigma$ arbitrary as in equations \eqref{Ricci_ell} and \eqref{K_ell}. Also, $\a \in \Real$ and $\mT_{ab}$ is so far an arbitrary symmetric tensor on $\Sigma$ with the above asymptotic expansion representing the first subleading corrections to the asymptotic value of $K_{[0]}$. As before, we will take the first condition \eqref{data_1} to be asymptotically equivalent to the solution \eqref{FG}:
\be\label{FG_ell}
h_{[0]ab}dx^{a}dx^{b}\, \sim\, {\ell^{2} \over z^{2}} \( dz^{2} + g_{ij}dx^{i}dx^{j} \)\ ,
\ee
with the Fefferman-Graham asymptotic expansion \eqref{FG_2}. Furthermore, since $h_{[0]} = \op(z^{-2})$ and we require the initial data set to be asymptotically hyperboloidal, we find that $\a > 0$.\\ 

The next step is to analyse the constraints on $\mT_{ab}$ imposed asymptotically by the initial data constraint equations \eqref{HamiltC} and \eqref{DiffeoC}. For our purposes, we will only need the constraints imposed on the leading order $\mT_{(0)}$. By using the asymptotic form \eqref{FG_ell} of $h_{[0]}$ we find that the constraint equations in a neighbourhood of the conformal boundary take the form:
\begin{align}
&\mT_{zz} + \Tr[g^{-1}\mT] + \qt\,z^{\a}\, \Big( 2\, \mT_{zz}\, \Tr[g^{-1}\mT] + \Tr^{2}[g^{-1}\mT] - 2\, \mT_{zi}\,g^{ij}\mT_{jz} - \Tr[g^{-1}\mT g^{-1}\mT]\, \Big) = 0\ ,\\[5pt]
&\na_{i} \( g^{ij} \mT_{jz} \)\ =\ {d \over z} \( \mT_{zz} + {\a - 1 \over d}\, \Tr[g^{-1}\mT] \) + \Tr[g^{-1}\mT ' ] - \hf\, \mT_{zz} \Tr[g^{-1}g'] - \hf\, \Tr[ g^{-1}g'g^{-1}\mT]\ , \\[15pt]
&\na_{j} \( g^{-1}\mT \)^{j}_{i} - \partial_{i}\, \Tr[g^{-1}\mT]\ =\ {d+1-\a \over z}\, \mT_{zi} - \mT_{zi}' - \hf\, \mT_{zi}\, \Tr[g^{-1}g'] + \partial_{i} \mT_{zz} \ ,
\end{align}
where $\na_{i}$ is the covariant derivative associated to $g_{ij}$ and $\mT':= \partial_{z}\mT$. From the leading order of the three constraint equations above we find the following three conditions for $\a \neq d+1$:
\be\label{cond_1}
\mT_{(0)zz} = 0\quad ,\quad \Tr[g_{(0)}^{-1}\mT_{(0)}] = 0\quad ,\quad \mT_{(0)zi} = 0 \ .
\ee
We also find the following additional condition for $\a = d$ from the first subleading order of the third constraint equation above:
\be\label{cond_2}
\cov_{j} \mT_{(0)i}^{j}\ =\ 0\ ,
\ee
where $\cov_{i}$ is the covariant derivative associated to $g_{(0)}$ and where the indices are raised with $g_{(0)}$. These four particular conditions will play an important role in the analysis that follows next. With these constraints identified, the approach we now take towards the initial value problem is to make a choice of lapse function and shift vector and solve asymptotically the evolution equations \eqref{evEq} and \eqref{dynEq} in powers of $z$ up to some desired order. If we find a solution to this problem up to some order in $z$, then such solution is the unique solution up to that order. We will analyse the three possible situations: when the power $\a$ is greater than, equal to or less than $d$ and we will find that the relevant case is when $\a = d$.\\

As in the previous sections, we begin by choosing to evolve the initial data in time in the gauge $(N=1,A^{a} = 0)$ in which the metric tensor \eqref{dev_met} of the Ricci-flat development in a neighbourhood of $\Sigma=\{\hat{t} = 0\}$ assumes the form \eqref{Gaussian_metric}:
\be
ds^{2}_{d+2} = -d\hat{t}^{\, 2} + h_{ab}dx^{a}dx^{b}\ .
\ee
The extrinsic curvature \eqref{evEq} on the surfaces of constant $\hat{t}$ in this gauge is given by $K_{ab} = \hf \partial_{\hat{t}}\, h_{ab}$. We then introduce a new coordinate $t$ as: $e^{t} := 1+\hat{t}/\ell$, in which $\Sigma = \{ t = 0\}$, and define $\tilde{h}_{ab} := \ell^{-2}e^{-2t} h_{ab}$. The development's metric therefore becomes:
\be\label{development}
ds^{2}_{d+2} = \ell^{2}e^{2t} \( -dt^{\, 2} + \tilde{h}_{ab}dx^{a}dx^{b} \) \ .
\ee
The extrinsic curvature on $\Sigma$ becomes:
\be
K_{[0]ab}\ =\ K_{ab}(t=0)\ =\ \left[ \ell\, e^{t} \( \tilde{h}_{ab} + \hf\, \partial_{t} \tilde{h}_{ab} \) \right]_{t=0} =\ \ell\,\tilde{h}_{ab}\Big|_{t=0} + {\ell \over 2}\, \partial_{t}\tilde{h}_{ab} \Big|_{t=0}\ .
\ee
In this way, the asymptotic initial conditions \eqref{data_1} and \eqref{data_2} in this coordinate system read respectively as:
\bea
&&\tilde{h}_{ab}dx^{a}dx^{b}\Big|_{t=0}\ =\ {1 \over z^{2}} \( dz^{2} + g_{ij}dx^{i}dx^{j} \)\ ,\label{data2_1}\\[5pt]
&&\partial_{t} \tilde{h}_{ab} \Big|_{t=0}\ =\ d\, z^{-2+\a}\, \mT_{ab}\ , \label{data2_2}
\eea
with $g_{ij}$ given by equation \eqref{FG_2}. We then solve asymptotically the dynamical equation obtained by replacing equation \eqref{evEq} in \eqref{dynEq} subject to the above initial value conditions. This equation reads as:
\be\label{dynamical}
2 \( R_{ab}[\tilde{h}] + d\, \tilde{h}_{ab} \) + \ddot{\tilde{h}}_{ab} + d\, \dot{\tilde{h}}_{ab} + \tilde{h}_{ab}\, \Tr[\tilde{h}^{-1}\dot{\tilde{h}}] + \hf\, \dot{\tilde{h}}_{ab}\, \Tr[\tilde{h}^{-1}\dot{\tilde{h}}] - \( \dot{\tilde{h}}\tilde{h}^{-1}\dot{\tilde{h}} \)_{ab}\ =\ 0\ ,
\ee
where $\dot{\tilde{h}} := \partial_{t}\tilde{h}$.\\

\subsection{Asymptotics and expectation values}

\n We begin by analysing the simplest case $\a > d$. Let one start with the following ansatz:
\begin{align}
&\tilde{h}_{ab}dx^{a}dx^{b}\ =\ {1 \over z^{2}} \( dz^{2} + \tilde{g}_{ij}dx^{i}dx^{j} \) + \op(z^{>-2+d}) \ , \label{ansatz}\\[5pt]
&\tilde{h}_{zi}\ =\ (1- e^{-dt})A_{[d]i}(z,x) + \op(z^{>-1+d})\ :\ A_{[d]i} = \op(z^{>-2+d})\ , \label{ansatz_zi} \\[5pt]
&\tilde{g}_{ij}(z,x)\ :=\ g_{(0)ij}(x) + z^{2}g_{(2)ij}(x) + ...  + z^{d} g_{(d)ij}(x) + z^{d}\log z\, \tilde{g}_{(d)ij}(x)\ ,\label{g_0}
\end{align}
where only even powers in $z$ arise below the order $z^{d}$ in the above finite series\footnote{ It should be emphasized that, unlike the expansion \eqref{FG_2}, the above expansion \eqref{g_0} is defined to be finite and to terminate at order $z^{d}$. 
}
and where each coefficient $g_{(n)}$ is defined to be equal to the coefficient $g_{(n)}$ in the Fefferman-Graham expansion \eqref{FG_2}. This ansatz satisfies the dynamical equation \eqref{dynamical} up to order $z^{-2+d}$ (see appendix \ref{appB1} for the complete treatment), and since $\a > d$, it also satisfies the initial conditions \eqref{data2_1} and \eqref{data2_2} up to order $z^{-2+d}$. In this way, the unique solution to this initial value problem must coincide with our ansatz up to order $z^{-2+d}$. Such solution receives corrections in all components at higher orders, but we will find that we do not need the asymptotic form of $\tilde{h}_{ab}$ beyond the order $z^{-2+d}$. Finally, we reinstate the lapse function $N(t)$ by redefining our time coordinate $t \to \log \int dt\, N(t)/\ell$ such that the development's metric becomes:
\be\label{development_beta}
ds^{2}_{d+2}\ =\ -N(t)^{2}dt^{2} + \beta(t)^{2} \tilde{h}_{ab}dx^{a}dx^{b}\ ,
\ee
with $\dot{\beta} := N(t)$ as before.\\

Given the above asymptotic solution for the Ricci-flat development (equation \eqref{development_beta} with \eqref{ansatz}) we then proceed as previously by renormalizing the gravitational action. If we regularize and evaluate the action \eqref{canonical_S} on-shell, we find that the divergences arise only down to order $\ve^{-d}$ as in \eqref{div_terms} and that these only involve the asymptotic form of $\tilde{h}_{ab}$ up to order $z^{-2+d}$. Hence, the counterterm action will be exactly the same as in \eqref{S_ren} because our solution is the same as in \eqref{metric} up to that order. We also find that the expectation values of the holographic stress tensors are equal to those found in \eqref{vev_T} because only the terms up to $z^{-2+d}$ in the asymptotic expansion of $\tilde{h}_{ab}$ survive under the limit $\epsilon \to 0$ taken in \eqref{def_T}. In this way, the case $\a > d$ does not yield new results.\\

The next step is to analyse the power $\a = d$. In this case, the four conditions \eqref{cond_1}--\eqref{cond_2} on the leading term $\mT_{(0)ab}$ hold. Let one start with the following ansatz:
\be\label{ansatz_1}
\tilde{h}_{ab}dx^{a}dx^{b} =\, {1 \over z^{2}}\, \Bigg[ dz^{2} + \bigg( \tilde{g}_{ij} + z^{d} \( \D_{ij} - \D_{[0]ij} \) \bigg) dx^{i}dx^{j} \Bigg]\, + \op(z^{>-2+d}) \ ,
\ee
with $\tilde{g}$ and $\tilde{h}_{zi}$ as defined in \eqref{ansatz_zi}--\eqref{g_0} and where: 
\be
\D_{ij} := \D_{ij}(t,x)\quad ,\quad \D_{[0]ij} := \D_{ij}(t=0,x)\ .
\ee
If we replace this ansatz in the dynamical equation \eqref{dynamical}, we find that this equation is solved up to order $z^{-2+d}$ if $\D_{ij}$ satisfies the second order differential equation in $t$:
\be
\ddot{\D}_{ij} + d\, \dot{\D}_{ij} = 0\quad \Rightarrow\quad \D_{ij}(t,x) = \D_{[0]ij}(x) + \( 1 - e^{-dt} \) \D_{[d]ij}(x)\ ,
\ee
subject to the conditions that the integration constant $\D_{[d]ij}$ be traceless and covariantly conserved with respect to $g_{(0)}$ (see appendix \ref{appB2}). In this way, the ansatz:
\begin{align}
&\tilde{h}_{ab}dx^{a}dx^{b} =\, {1 \over z^{2}}\, \Bigg[ dz^{2} + \bigg( \tilde{g}_{ij} + z^{d} \( 1 - e^{-dt} \) \D_{[d]ij}(x) \bigg) dx^{i}dx^{j} \Bigg]\, + \op(z^{>-2+d}) \ , \label{ansatz_2}\\[5pt]
&\Tr[g_{(0)}^{-1}\D_{[d]}] = 0\quad ,\quad \cov_{j} \D^{j}_{[d]i} = 0 \label{ansatz_2-2}\ ,
\end{align}
solves the dynamical equation up to order $z^{-2+d}$. If we then compute the initial values $\tilde{h}_{ab}(t=0)$ and $\partial_{t}\tilde{h}_{ab}(t=0)$, we find that these coincide with the initial conditions \eqref{data2_1} and \eqref{data2_2} up to order $z^{-2+d}$ (recall that $\mT_{(0)za} = 0$) if we identify $\D_{[d]ij}$ with $\mT_{(0)ij}$:
\be\label{identification}
\D_{[d]ij}\ =\ \mT_{(0)ij}\ .
\ee
Recall that the leading order $\mT_{(0)ij}$ is also traceless and covariantly conserved. In this way, with the above identification, we find that our ansatz \eqref{ansatz_2} solves the initial value problem up to order $z^{-2+d}$ and therefore the unique solution to this problem must coincide with our ansatz up to this order. Finally, we reinstate the lapse function $N(t)$ as in \eqref{development_beta} to obtain:
\be
ds^{2}_{d+2} = -N(t)^{2} dt^{2} + \beta(t)^{2} \( {dz^{2} \over z^{2}} + {1 \over z^{2}} \Big[ \tilde{g}_{ij} + z^{d} \( 1 - \big(\beta(t)/\ell \big)^{-d} \) \mT_{(0)ij} \Big] dx^{i}dx^{j} + \op(z^{>-2+d}) \) ,
\ee
with $\dot{\beta} = N(t)$. Given the above asymptotic solution we then proceed to compute the holographic stress tensors. The counterterm action will again be the same as in \eqref{S_ren} because our asymptotic solution for the Ricci-flat development only differs from the previous solution \eqref{metric} at order $z^{-2+d}$. The divergences of the on-shell action involve the asymptotic form of the Ricci-flat metric below this order and the trace of $\tilde{h}_{ij}$ with respect to $g_{(0)}$ at order $z^{-2+d}$. However, since the trace of $\mT_{(0)ij}$ vanishes, the trace of $\tilde{h}_{ij}$ is the same as that computed from \eqref{metric} up to order $z^{-2+d}$ and hence we find the same divergent terms as previously, as well as the same counterterms. If we then compute the holographic stress tensors according to formula \eqref{def_T}, we find them to be of the same form as in \eqref{vev_T} plus an additional contribution:
\be\label{vev_T_gen}
\< T_{ij} \>\ =\ {d\, \beta(t)^{d} \over 2\kappa_{d+2}^{2}}\, \Big( g_{(d)ij} + \mT_{(0)ij} + X_{ij}[g_{(0)}] \Big) - {d\, \ell^{d} \over 2\kappa_{d+2}^{2}}\, \mT_{(0)ij}\ .
\ee
Notice that we recover the result \eqref{vev_T} for the initial hypersurface $\Sigma=\{\beta(t)=\ell\}$. The diffeomorphism Ward identities are still as in equation \eqref{diffeo_ward} and the trace Ward identities
are also as in \eqref{trace_ward_odd} for odd values of $d$. For even values the trace depends on $d$. For $d+2=4$, we have:
\bea
\< T_{ij} \> &=& {\beta(t)^{2} \over \kappa_{4}^{2}} \( g_{(2)ij} + \mT_{(0)ij} -g_{(0)ij}\Tr[g_{(0)}^{-1}g_{(2)}] \) - {\ell^{2} \over \kappa_{4}^{2}}\, \mT_{(0)ij}\ , \label{vev_T_2}\\
\< T^{i}_{i} \> &=& {\beta(t)^{2} \over 2\kappa^{2}_{4}}\, R[g_{(0)}]\ =\ {c \over 24\pi}\, R[g_{(0)}]\ .
\eea
For small and large values of the central charge, we find the asymptotic behaviours:
\bea
\< T_{ij} \>\ =\ 
\begin{cases}
 - {\ell^{2} \over \kappa_{4}^{2}}\, \mT_{(0)ij} + \op(\beta^{2})\quad :\ \beta \ll 1\ ,\\[5pt]
{\beta^{2} \over \kappa_{4}^{2}} \( g_{(2)ij} + \mT_{(0)ij} -g_{(0)ij}\Tr[g_{(0)}^{-1}g_{(2)}] \) + \op(\beta^{0})\quad\ :\ \beta \gg 1\ .
\end{cases}
\eea
Different field theories in the family, identified by their central charges, have therefore different expectation values of the stress tensors and, in order to reconstruct the bulk spacetime metric from CFT data in this case, we need such expectation values in the regimes of large and small central charges.\\

The last case to be analysed is for $\a < d$. In this case, only the three conditions \eqref{cond_1} on the leading term $\mT_{(0)ab}$ hold. The strategy is again to start from an ansatz as in \eqref{ansatz_1}, where the power $z^{d}$ is now replaced by $z^{\a}$, and find the differential equation and constraints that $\D_{ij}$ needs to satisfy in order for the dynamical equation to be solved up to order $z^{-2+\a}$. We then compute the initial values $\tilde{h}(t=0)$ and $\partial_{t}\tilde{h}(t=0)$ and find the relations between the integration constants in the solution for $\D_{ij}$ and the leading order $\mT_{(0)ij}$ in order for the initial conditions to be satisfied up to order $z^{-2+\a}$. The unique solution to this initial value problem must then coincide with our refined ansatz up to order $z^{-2+\a}$. The next step is to attempt to renormalize the gravitational action \eqref{canonical_S}. If we regularize it as previously and evaluate it on-shell, we find that the divergences involve the asymptotic form of the Ricci-flat metric up to order $z^{-2+d}$ as before. However, we have just deduced that this asymptotic form involves terms already at order $z^{-2+\a}$ (and beyond) that are not pure functionals of $g_{(0)ij}$. Such terms are rather functionals of the unspecified leading and subleading orders $\mT_{(n)ab}$ in the asymptotic expansion of $\mT_{ab}$. In this way, the divergences will be functionals of the undetermined terms $g_{(0)}$ and $\mT_{(n)}$. This implies that we cannot rewrite the divergences covariantly in terms of the induced metric $\g_{ij}=h_{ij}$ (or $\g_{AB}$) as in equation \eqref{pre_CT}. We can invert the asymptotic series for $\tilde{h}_{ij}$ as before in order to rewrite $g_{(0)ij}$ covariantly in terms of $\g_{ij}$, but now only up to order $z^{-2+\a}$, which is below $z^{-2+d}$. Furthermore, the divergences involving the terms $\mT_{(n)}$ cannot be rewritten in a covariant fashion because the asymptotic expansion for $\mT_{ab}$ is undetermined and unrelated to $\g_{ij}$ and hence cannot be inverted in order to express the terms $\mT_{(n)}$ as functionals of $\g_{ij}$. In the language of holographic renormalization, this type of divergent terms are said to be non-local in the sources. We therefore find that if the exponent $\a < d$, the Ricci-flat development is non-renormalizable holographically in the sense that the gravitational action for the development cannot be renormalized with local covariant counterterms and the only perturbations of the initial data $K_{[0]}$ away from the hyperboloidal one that result in renormalizable developments must occur at order equal to or greater than $z^{-2+d}$.

\section{Conclusions}

\qquad In this work we have found that there exists a natural generalisation of Minkowski space for which a holographic description in terms of a family of conformal field theories on a codimension two manifold seems possible. This was achieved for pure bulk gravity by showing that the asymptotics of every Ricci-flat spacetime admitting a hypersurface that is both locally hyperbolic and of constant mean curvature sufficiently close to null infinity can be reconstructed near such surface from a family of CFT data. We have deduced the most general asymptotics of this class of spacetimes near any codimension two manifold at null infinity, renormalized holographically the gravitational action with vanishing cosmological constant, reproduced the Ward identities and vacuum expectation values of a family of CFT stress tensors and mapped the bulk data necessary to the reconstruction of the spacetime asymptotics to the sources and one-point correlators in this family of field theories. In even dimensions, from the holographic Weyl anomalies of each CFT we have argued that locality in the bulk time direction is recovered only if there exists a notion of locality in the spectrum of central charges in this family. This feature requires a deeper understanding of the interrelationship between the different field theories required by the duality and we have suggested that one direction to this end is the study of the symmetry group in the family associated to metric-preserving bulk diffeomorphisms that involve the time coordinate. 

Another aspect that should be understood further within this framework is the case of asymptotically flat black holes such as the Schwarzschild spacetime. In a very first approach to this problem, it seems that, at least in four bulk dimensions and at the level of the one-point correlators, the family of dual field theories does not distinguish between Minkowski space and the Schwarzschild solution: since the latter approaches the former asymptotically, then Schwarzschild can be written in the gauge \eqref{metric} in a sufficiently small neighbourhood of $\del \mH$. Then, by comparison to Minkowski \eqref{mink_II}, corrections must occur at order higher that $z^{4}$ in $g_{ij}$. However, in four bulk dimensions, the expectation values of the CFT stress tensors only care about the asymptotics up to order $z^{2}$ which is the same between Schwarzschild and Minkowski. A solution to this issue may be that, as one moves sufficiently away from $\del \mH$, the gauge \eqref{metric} is necessarily broken for Schwarzschild and cross terms between the spatial, radial and time components of the metric start to arise at a sufficiently low order in $z$, even though corrections of $g_{ij}$ only happen at order $z^{>4}$. Such gauge-spoiling terms would require a different holographic renormalization procedure and result in expectation values that would differ from those obtained in \eqref{vev_T} and more generally in \eqref{vev_T_gen}.

Finally, the holographic reconstruction of bulk matter in our class of spacetimes from CFT data is an aspect to be pursued in future work. As argued at the end of section \ref{holog_energy-tensor}, a first contact with the subject seems to reveal that not all solutions of the bulk matter equations admit a one-to-one correspondence with a single operator per field theory as in the case of the spacetime metric and the stress tensors, but there should nevertheless exist a subclass of solutions that suffices to capture the dynamics of gauge-invariant operators in the family of field theories.\\

\section*{Acknowledgments}

I would like to thank Marika Taylor, Kostas Skenderis, Sergey Solodukhin and Jan de Boer for valuable discussions and comments on the manuscript. I am also indebted to Ioannis Papadimitriou for insightful conversations at different stages of this work, in particular on the subject of holographic renormalization. I gratefully acknowledge support from the Funda\c{c}\~{a}o para a Ci\^{e}ncia e Tecnologia (FCT, Portugal) via the grant SFRH/BD/43182/2008.

\vspace{30pt}

\appendix

\section*{Appendix}
\addcontentsline{toc}{section}{Appendix:}

\vspace{4pt}

\section{Conformal compactness and asymptotia}\label{appA}

In this appendix we deduce a few implications of conformal compactness that are relevant to our work. We begin with the standard definition.\\

A manifold $(\mM,G)$ is defined to be $C^{n\geq 0}$ conformally compact if there exists an {\it asymptote} $(\tilde{\mM},\tilde{G},\r)$ consisting of a defining function $\r(x) \in [0,+\infty[$ and a manifold-with-boundary $(\tilde{\mM},\tilde{G})$ with boundary $\del \tilde{\mM}$ satisfying the following properties \cite{penrose,geroch,Anderson:2004yi}:

\begin{itemize}
\item[1)] $\mM= \text{int } \tilde{\mM} = \{p \in \tilde{\mM} : \exists\ \text{open set}\ p \ni U \subset \tilde{\mM} \} \ , $
\item[2)] $\tilde{G}_{\mu\nu} = \r^{2}(x)\, G_{\mu\nu}\ $ : $\ \tilde{\mM} = \{\r \geq 0\}$ , $\del \tilde{\mM} = \{\r = 0\}\ ,$
\item[3)] $d\r \neq 0$ on $\del \tilde{\mM}$ ,
\end{itemize}

\n with $\r(x)$ of class $C^{\infty}$ and $\tilde{G}$ non-degenerate and of class $C^{n \geq 0}$ in $\tilde{\mM}$. The region $\{\r=0\}$ of $\tilde{\mM}$ is referred to as the conformal boundary of $\mM$ and $\tilde{\mM}$ as the conformal embedding.\\

We also define an asymptotically locally flat (AlF) space as any conformally compact, asymptotically Ricci-flat Lorentzian manifold. This definition essentially coincides with that of asymptotic flatness at null infinity \cite{wald} if we relax any conditions on the topology of the conformal boundary.\\

We are now interested  in showing that a conformally compact, asymptotically Einstein Riemannian manifold of negative scalar curvature is asymptotically hyperbolic up to global identifications and also that the boundary of an AlF space is null. In order to do so, we need the following result.

\paragraph{Proposition}\label{Prop_1}

Let $(\mM,G)$ be a conformally compact manifold with an asymptote $(\tilde{\mM},\tilde{G},\rho)$. Then in the limit $\rho \to 0$, the Riemann tensor behaves asymptotically as:
\be\label{asymptR}
R_{abcd} = - |\tilde{\na}\r|^{2} \left( G_{ac}G_{bd} -G_{ad}G_{bc} \right) + \op(\r^{>-4}) \quad ,
\ee
where: $|\tilde{\na}\r|^{2} := \tilde{G}^{ab}\partial_{a}\r\partial_{b}\r$ and where $\op(\r^{>-4})$ denotes terms that diverge slower than $\r^{-4}$.

\paragraph{Proof}

By using the transformation law of the Riemann tensor under a conformal transformation, the Riemann of $G$ and that of $\tilde{G}$ are related as:
\be\label{RTransf}
R_{abcd} = \r^{-2}\tilde{R}_{abcd} + \left( \r^{-3} \tilde{G} \circ \tilde{\na} \tilde{\na} \r -\hf \r^{-4}|\tilde{\na}\r|^{2} \tilde{G} \circ \tilde{G} \right)_{abcd} \quad ,
\ee
where: $ \left( A \circ B \right)_{abcd} :=  A_{ac}B_{bd} - A_{ad}B_{bc} + B_{ac}A_{bd} - B_{ad}A_{bc}$, and where $\tilde{\na}_{a}$ is the covariant derivative with respect to $\tilde{G}$. Then, from the third condition in the definition of conformal compactness, we can introduce the defining function $\r$ as a coordinate in the neighbourhood of $\r=0$. In this way, since $\tilde{G}$ is at least $C^{0}$ in $\tilde{\mM}$ and the Riemann tensor of $\tilde{G}$ contains at most second derivatives of $\tilde{G}$ with respect to $\r$, then if $\tilde{R}_{abcd}$ diverges as $\r \to 0$, it must do so slower than $\r^{-2}$. Also, the term $\tilde{\na}_{a}\tilde{\na}_{b}\r$ contains at most first derivatives of $\tilde{G}$ with respect to $\r$ and hence it must diverge slower than $\r^{-1}$. Hence, the first two terms in \eqref{RTransf} diverge slower than $\r^{-4}$ and thus we find:
\bea
R_{abcd} &=& - \r^{-4} |\tilde{\na}\r|^{2} \left( \tilde{G}_{ac}\tilde{G}_{bd} - \tilde{G}_{ad}\tilde{G}_{bc} \right) + \op(\r^{>-4})\nn\\
&=& - |\tilde{\na}\r|^{2} \left( G_{ac}G_{bd} - G_{ad}G_{bc} \right) + \op(\r^{>-4}) \quad . 
\eea\\[5pt]
Now, from equation \eqref{asymptR}, the Ricci tensor of $G$ behaves asymptotically as:
\be\label{asymptRicci}
R_{ab}\ =\ -d\, \r^{-2} |\tilde{\na}\r|^{2} \tilde{G}_{ab} + \op(\r^{>-2})\ \sim -d\, |\tilde{\na}\r|^{2} G_{ab}\ ,
\ee
where $d+1$ is the dimension of $\mM$. 

If $(\mM,G)$ is in particular asymptotically Einstein of negative scalar curvature, then from the above we find that $|\tilde{\na}\r|^{2}$ must be a positive constant by definition. From equation \eqref{asymptR} it then follows that the Riemann tensor is asymptotically equal to that of Anti-de Sitter or of the hyperbolic space for $(\mM,G)$ Lorentzian or Riemannian, respectively. In the latter case, $(\mM,G)$ is therefore asymptotically locally isometric to the hyperbolic space. 

If $(\mM,G)$ is, on the other hand, asymptotically Ricci-flat, then from \eqref{asymptRicci} we find that $|\tilde{\na}\r|^{2}$ must vanish asymptotically as $\rho \to 0$. Since $d\rho \neq 0$ by definition, this implies that the conformal boundary $\{\rho = 0\}$ of $\mM$ is null.\\

Finally, we show that the Ricci scalar of a conformally compact Riemannian manifold $(\mM,G)$ of dimension $d+1 > 1$ cannot vanish asymptotically. From equation \eqref{asymptR}, the Ricci scalar of $G$ behaves asymptotically as:
\be\label{SCurv}
R = - d(d+1) |\tilde{\na}\r|^{2} + \op(\r^{>0}) \quad .
\ee
If $(\mM,G)$ is Riemannian, one can write without loss of generality the positive-definite metric tensor $G_{ab}$ near $\r=0$ as:
\bea\label{ADM-G}
ds^{2} &=& G_{ab}dx^{a}dx^{b}\nn\\[5pt]
&=& M^{2} d\r^{2} + \g_{ij} \left( dx^{i} + B^{i}d\r \right) \left( dx^{j} + B^{j} d\r \right)\nn\\[5pt]
&=& \r^{-2} \left( \tilde{M}^{2} d\r^{2} + \tilde{\g}_{ij} \left( dx^{i} + B^{i}d\r \right) \left( dx^{j} + B^{j} d\r \right) \right)\nn\\[5pt]
&=& \r^{-2} \tilde{G}_{ab}dx^{a}dx^{b}\quad .
\eea
From this decomposition, we find:
\be
|\tilde{\na} \r|^{2} = \tilde{G}^{\r\r} = \tilde{M}^{-2} \quad.
\ee
Since $\tilde{G}_{ab}$ is at least of class $C^{0}$ and $\tilde{G}_{\r\r} = \tilde{M}^{2} + \tilde{\g}_{ij}B^{i}B^{j}$, then $\tilde{M}^{-2}$ must be supported in $\tilde{\mM}$, otherwise $\tilde{G}_{\r\r}$ would not be $C^{0}$. Notice that $\tilde{\g}_{ij}$, being positive-definite, prevents the term $\tilde{\g}_{ij}B^{i}B^{j}$ from cancelling $\tilde{M}^{2}$ in $\tilde{G}_{\r\r}$.
In this way, $|\tilde{\na}\r|^{2}$ is non-vanishing and thus the Ricci scalar cannot vanish asymptotically.

\section{Asymptotic solutions of the dynamical equation}

\subsection{Solution for $\a>d$}\label{appB1}

In this section we show that the ansatz \eqref{ansatz}--\eqref{g_0} solves the dynamical equation \eqref{dynamical} up to order $z^{-2+d}$. Let us begin by performing an ADM decomposition of $\tilde{h}_{ab}$ with respect to surfaces of constant $z$:
\be\label{B-ADM}
\tilde{h}_{ab}dx^{a}dx^{b} = M^{2}dz^{2} + \g_{ij} ( dx^{i} + A^{i}dz ) ( dx^{j} + A^{j}dz )\ ,
\ee
with unit normal $n_{a} = M\partial_{a}z$. The inverse of $\tilde{h}_{ab}$ is given by:
\bea
\tilde{h}^{ab} = 
\begin{pmatrix} 
M^{-2} & -M^{-2} A^{i}\\
-M^{-2} A^{j}\quad & \g^{ij} + M^{-2} A^{i}A^{j}
\end{pmatrix}\ .
\eea
We then start from the following ansatz:
\bea\label{B-ansatz}
\begin{cases}
M\ =\ z^{-1} + \op(z^{>-1+d})\quad , \\[5pt]
A_{i}\ =\ (1-e^{-dt})A_{[d]i}(z,x) + \op(z^{>-1+d})\quad :\quad A_{[d]i} = \op(z^{>-2+d})\quad , \\[5pt]
\g_{ij}\ =\ z^{-2} g_{ij}\quad :\quad g_{ij}(t,z,x) = \tilde{g}_{ij}(z,x) + \op(z^{>d})\ ,\ \tilde{g}_{ij} = \op(z^{0})\quad , 
\end{cases}
\eea
such that:
\be
\tilde{h}_{ab}dx^{a}dx^{b} = {1 \over z^{2}} \( dz^{2} + \tilde{g}_{ij}dx^{i}dx^{j} \) + \op(z^{>-2+d})\ .
\ee
The reason for the above dependence of the shift $A_{i} = \g_{ij}A^{j}$ on $t$ at orders between $\op(z^{>-2+d})$ and $\op(z^{-1+d})$ will become clear later. If we replace this ansatz in the dynamical equation:
\be\label{B-dynamical}
2 \( R_{ab}[\tilde{h}] + d\, \tilde{h}_{ab} \) + \ddot{\tilde{h}}_{ab} + d\, \dot{\tilde{h}}_{ab} + \tilde{h}_{ab}\, \Tr[\tilde{h}^{-1}\dot{\tilde{h}}] + \hf\, \dot{\tilde{h}}_{ab}\, \Tr[\tilde{h}^{-1}\dot{\tilde{h}}] - \( \dot{\tilde{h}}\tilde{h}^{-1}\dot{\tilde{h}} \)_{ab}\ =\ 0\ ,
\ee
we find:
\bea
&&R_{ij}[\tilde{h}] + d\, \tilde{h}_{ij} + \op(z^{>-2+d})\ =\ 0\quad , \label{B-ij}\\[10pt]
&&n^{a}n^{b}R_{ab}[\tilde{h}] + d + \op(z^{>d})\ =\ 0\quad , \label{B-zz}\\[10pt]
&&n^{a}R_{ai}[\tilde{h}] + \hf\, n^{a} \( \ddot{\tilde{h}}_{ai} + d\, \dot{\tilde{h}}_{ai} \) + \op(z^{>-1+2d})\ =\ 0\quad . \label{B-zi}
\eea
We begin by analysing the spatial components $(i,j)$. In order to do so, we need the Gauss-Codazzi identity:
\bea
K_{ij}' &=& \Lie_{A}K_{ij} - D_{i}D_{j}M + M\, \bigg( R_{ij}[\g] + 2 (K \g^{-1}K)_{ij} - K_{ij} \Tr[\g^{-1}K] - R_{ij}[\tilde{h}] \bigg)\ ,\\[10pt]
K_{ij} &=& {1 \over 2M} \( \g_{ij}' - \Lie_{A}\g_{ij} \)\ , \label{B-Kij}
\eea
where prime denotes differentiation with respect to $z$ and $D_{i}\g_{jk} := 0$. If we use our ansatz in \eqref{B-Kij}, we find:
\be\label{B-Kij_2}
K_{ij} = -z^{-2} g_{ij} + \hf z^{-1}g_{ij}' + \op(z^{>-2+d})\ .
\ee
With this result, the Gauss-Codazzi identity becomes:
\bea\label{B-GC}
2 \( R_{ij}[\tilde{h}] + {d \over z^{2}}\, g_{ij} \) &=& 2 R_{ij}[g] - g_{ij}'' + {d-1 \over z}\, g_{ij}' + {1 \over z}\, g_{ij} \Tr[g^{-1}g'] - \hf\, g_{ij}' \Tr[g^{-1}g'] \nn\\
&& + \( g' g^{-1} g' \)_{ij} + \op(z^{>-2+d})\ .
\eea
If we then replace this identity in \eqref{B-ij}, we obtain:
\be\label{B-FG}
2 R_{ij}[g] - g_{ij}'' + {d-1 \over z}\, g_{ij}' + {1 \over z}\, g_{ij} \Tr[g^{-1}g'] - \hf\, g_{ij}' \Tr[g^{-1}g'] + \( g' g^{-1} g' \)_{ij} + \op(z^{>-2+d}) = 0\ .
\ee
The above is the Fefferman-Graham (FG) equation plus additional contributions at order $z^{>-2+d}$ and therefore is solved by the FG solution up to order $z^{d}$ in $g_{ij}$:
\be\label{B-FG_2}
g_{ij} = \underbrace{g_{(0)ij} + z^{2} g_{(2)ij} + ... + z^{d}\log z\, \tilde{g}_{(d)ij} + z^{d} g_{(d)ij}}_{\tilde{g}} + \op(z^{>d})\ ,
\ee
with $g_{(n)}$ the FG coefficients. In this way, our ansatz \eqref{ansatz}--\eqref{g_0} solves the spatial components of the dynamical equation up to order $z^{-2+d}$. The same procedure can be applied to the remaining components \eqref{B-zz} and \eqref{B-zi}. For the former, we need the Gauss-Codazzi identity:
\be
-R[\g] + \Tr^{2}[\g^{-1}K] - \Tr[\g^{-1}K\g^{-1}K]\ =\ n^{a}n^{b}R_{ab}[\tilde{h}] - \g^{ij}R_{ij}[\tilde{h}]\ ,
\ee
where $R[\g]$ is the Ricci-scalar of $\g_{ij}$ and $n^{a} = \tilde{h}^{ab}n_{b}\ :\ n_{b} = M\partial_{b}z$. If we use our ansatz \eqref{B-ansatz} in this identity, together with the expression \eqref{B-Kij_2} for $K_{ij}$ and the previous Gauss-Codazzi identity \eqref{B-GC}, we find:
\be
n^{a}n^{b}R_{ab}[\tilde{h}] = -d - {z^{2} \over 2} \( \Tr[g^{-1}g''] - {1 \over z}\, \Tr[g^{-1}g'] - \hf\, \Tr[g^{-1}g'g^{-1}g'] \) + \op(z^{>d})\ .
\ee
If we then replace this identity in the equation \eqref{B-zz}, we obtain:
\be\label{B-FG_Tr}
\Tr[g^{-1}g'']-{1 \over z}\, \Tr[g^{-1}g'] - \hf\, \Tr[g^{-1}g'g^{-1}g'] + \op(z^{>-2+d}) = 0\ .
\ee
The above is the FG trace equation plus additional contributions at order $z^{>-2+d}$ and is solved by the FG solution up to order $z^{d}$ in $g_{ij}$. Hence, the components \eqref{B-zz} of the dynamical equation are also solved by our ansatz \eqref{ansatz}--\eqref{g_0} up to order $z^{-2+d}$ in $\tilde{h}_{ab}$. Finally, for the remaining components \eqref{B-zi} we need the last Gauss-Codazzi identity:
\be
D_{j} \( \g^{-1}K \)^{j}_{\, i} - D_{i} \Tr[\g^{-1}K] = n^{a}R_{ai}[\tilde{h}]\ .
\ee
If we use again the ansatz \eqref{B-ansatz} in this identity, we find:
\be
n^{a}R_{ai}[\tilde{h}] = {z \over 2} \( \na_{j} \( g^{-1}g' \)^{j}_{\, i} - \na_{i} \Tr[g^{-1}g'] \) + \op(z^{>d})\ ,
\ee
where $\na_{i}g_{jk} := 0$. By replacing this identity in \eqref{B-zi}, we obtain:
\be\label{B-FG_Div_0}
\na_{j} \( g^{-1}g' \)^{j}_{\, i} - \na_{i} \Tr[g^{-1}g'] + z^{-1} n^{a} \( \ddot{\tilde{h}}_{ai} + d\, \dot{\tilde{h}}_{ai} \) + \op(z^{>-1+d}) = 0 \ .
\ee
Now notice that $n = z \partial_{z} + \op(z^{>1+d})$ and hence that:
\be
z^{-1}n^{a} \( \ddot{\tilde{h}}_{ai} + d\, \dot{\tilde{h}}_{ai} \) = \ddot{\tilde{h}}_{zi} + d\, \dot{\tilde{h}}_{zi} + \op(z^{>-2+2d})\ .
\ee
Since $\tilde{h}_{zi} = A_{i}$, we obtain:\footnote{ In general we can have $A_{i} = A_{[0]i}(z,x) - e^{-dt}A_{[d]i}(z,x) + \op(z^{>-1+d})\ :\ A_{[0,d]i} = \op(z^{>-2+d})$, but we set $A_{[0]} = A_{[d]}$ so that our ansatz satisfies the initial condition \eqref{data2_1}.}
\be
z^{-1}n^{a} \( \ddot{\tilde{h}}_{ai} + d\, \dot{\tilde{h}}_{ai} \) = \op(z^{>-1+d})\ ,
\ee
and hence equation \eqref{B-FG_Div_0} becomes:
\be\label{B-FG_Div}
\na_{j} \( g^{-1}g' \)^{j}_{\, i} - \na_{i} \Tr[g^{-1}g'] + \op(z^{>-1+d}) = 0 \ .
\ee
This last equation is the FG divergence equation plus additional contributions at order $z^{>-1+d}$ and is also solved by the FG solution up to order $z^{d}$ in $g_{ij}$. Hence, we conclude that our ansatz \eqref{ansatz}--\eqref{g_0} solves the dynamical equation \eqref{dynamical} up to order $z^{-2+d}$ in \nolinebreak $\tilde{h}_{ab}$.

\subsection{Solution for $\a=d$}\label{appB2}

We now show that the ansatz \eqref{ansatz_2}--\eqref{ansatz_2-2} solves the dynamical equation \eqref{dynamical} up to order $z^{-2+d}$. We perform again the ADM decomposition \eqref{B-ADM} of $\tilde{h}_{ab}$ and start from the ansatz \eqref{B-ansatz}, but now do not impose the dependence of $g_{ij}$ on $t$ to arise at order $z^{>d}$:
\bea\label{B2-ansatz}
\begin{cases}
M\ =\ z^{-1} + \op(z^{>-1+d})\quad , \\[5pt]
A_{i}\ =\ (1-e^{-dt})A_{[d]i}(z,x) + \op(z^{>-1+d})\quad :\quad A_{[d]i} = \op(z^{>-2+d})\quad , \\[5pt]
\g_{ij} = z^{-2} g_{ij}(t,z,x)\ :\ g_{ij} = \op(z^{0})\ . 
\end{cases}
\eea
If we replace such ansatz in the dynamical equation \eqref{B-dynamical} and use again the Gauss-Codazzi identities as in the previous section, we obtain:
\begin{align}
& -\ddot{g}_{ij} - d\, \dot{g}_{ij} - g_{ij} \Tr[g^{-1}\dot{g}] - \hf\, \dot{g}_{ij} \Tr[g^{-1}\dot{g}] + \( \dot{g}g^{-1}\dot{g} \)_{ij} = \nn \\
& z^{2} \( 2 R_{ij}[g] - g_{ij}'' + {d-1 \over z}\, g_{ij}' + {1 \over z}\, g_{ij}\, \Tr[g^{-1}g'] - \hf\, g_{ij}' \Tr[g^{-1}g'] + \( g' g^{-1} g' \)_{ij} \) + \op(z^{>d}) \ , \label{B2-ij}\\[10pt]
& \Tr[g^{-1}\dot{g}]\ =\ z^{2} \( \Tr[g^{-1}g''] - {1 \over z}\, \Tr[g^{-1}g'] - \hf\, \Tr[g^{-1}g'g^{-1}g'] \) + \op(z^{>d}) \ , \label{B2-zz}\\[10pt]
& \na_{j} \( g^{-1}g' \)^{j}_{i} - \na_{i} \Tr[g^{-1}g'] + + \op(z^{>-1+d})\ =\ 0\ . \label{B2-zi}
\end{align}
These are the analogues of equations \eqref{B-FG}, \eqref{B-FG_Tr} and \eqref{B-FG_Div}. We then seek for a solution of the form:
\bea
g_{ij}(t,z,x) &=& g_{(0)ij}(x) + z^{2}g_{(2)ij}(x) + ... + z^{d}\, \Big( g_{(d)ij}(x) + \D_{ij}(t,x) - \D_{[0]ij}(x) \Big) \nn\\[5pt]
&& +\, z^{d}\log z\, \tilde{g}_{(d)ij}(x) + \op(z^{>d})\ ,
\eea
where only even powers in $z$ arise below the order $z^{d}$ and where each coefficient $g_{(n)}$ is defined to be the FG coefficient in the solution \eqref{B-FG_2}. Also, $\D_{[0]} := \D(t=0)$. If we replace for $g_{ij}$ in \eqref{B2-ij}--\eqref{B2-zi}, we find that the equations are satisfied up to order $z^{d}$ in $g_{ij}$ if $\D_{ij}$ obeys the equations:
\bea
&&\ddot{\D}_{ij} + d \dot{\D}_{ij} + g_{(0)ij} \Tr[g^{-1}_{(0)}\dot{\D}] + d\, g_{(0)ij} \Tr[g^{-1}_{(0)}\( \D - \D_{[0]} \) ] = 0\ ,\label{B-Dij}\\[5pt]
&&\Tr[g_{(0)}^{-1}\dot{\D}] - d(d-2) \Tr[g_{(0)}^{-1}\( \D - \D_{[0]} \) ] = 0\ , \label{B-D_Tr}\\[5pt]
&&\cov_{j} \(g_{(0)}^{-1} \( \D - \D_{[0]} \) \)^{j}_{\, i} - \partial_{i} \Tr[g_{(0)}^{-1}\( \D - \D_{[0]} \)] = 0\ , \label{B-D_div}
\eea
where $\cov_{j} g_{(0)ik} := 0$. Suppose that $\Tr[g_{(0)}^{-1}\dot{\D}] = 0$. Then, equation \eqref{B-D_Tr} is solved (recall that $\dot{g}_{(0)}=0$) and equation \eqref{B-Dij} becomes:
\be
\ddot{\D}_{ij} + d\, \dot{\D}_{ij} = 0\quad \Rightarrow\quad \D_{ij}(t,x) = \D_{[0]ij}(x) + \( 1 - e^{-dt} \) \D_{[d]ij}(x)\ .
\ee
If we insert this solution in the remaining equation \eqref{B-D_div}, we obtain:
\be
\cov_{j} \(  g_{(0)}^{-1} \D_{[d]} \)^{j}_{\, i} = 0\ .
\ee
In this way, the ansatz:
\bea
&&g_{ij}(t,z,x)\ =\ g_{(0)ij}(x) + z^{2}g_{(2)ij}(x) + ... + z^{d}\, \Big( g_{(d)ij}(x) + \( 1 - e^{-dt} \) \D_{[d]ij}(x) \Big) \nn\\[5pt]
&&\phantom{g_{ij}(t,z,x)\ } +\, z^{d}\log z\, \tilde{g}_{(d)ij}(x) + \op(z^{>d})\ : \nn\\[5pt]
&&\Tr[g_{(0)}^{-1}\D_{[d]}] = 0\ ,\ \cov_{j} \D^{j}_{[d]i} = 0\ ,
\eea
is a solution to the equations \eqref{B2-ij}--\eqref{B2-zi} up to order $z^{d}$ in $g_{ij}$ and therefore our ansatz \eqref{ansatz_2}--\eqref{ansatz_2-2}, or \eqref{B2-ansatz} with $g_{ij}$ as above, solves the dynamical equation \eqref{dynamical} up to order $z^{-2+d}$ in $\tilde{h}_{ab}$.

\newpage

\bibliographystyle{utphys}
\bibliography{biblio}

\end{document}